\documentclass[11pt]{article}
\usepackage{amsmath,amsthm,amsfonts,amssymb,amsopn,amscd}     
\usepackage{epsfig}
\usepackage{graphicx,tikz}
\usetikzlibrary{arrows}
\usepackage{setspace}
\usepackage{stmaryrd}
\newcommand{\tikzmatht}[2][0.42]
{\vcenter{\hbox{\begin{tikzpicture}[scale=#1]\footnotesize #2
				 \end{tikzpicture}}}
}
\DeclareMathOperator{\Tr}{Tr}

\def\Ad{{\mathrm{Ad}}}

\def\dim{{\mathrm{dim}}}
\def\End{{\mathrm{End}}}
\def\Hom{{\mathrm{Hom}}}
\def\id{{\mathrm{id}}}

\def\Mat{\mathrm{Mat}}

\newtheorem{theorem}{Theorem}[section]
\newtheorem{lemma}[theorem]{Lemma}

\newtheorem{corollary}[theorem]{Corollary}

\newtheorem{proposition}[theorem]{Proposition}

\theoremstyle{definition} 
\newtheorem{definition}[theorem]{Definition}
\newtheorem{remark}[theorem]{Remark}

\def\Vir{{\mathrm {Vir}}}

\def\RR{{\mathbb R}}
\def\CC{{\mathbb C}}

\def\MM{{\mathbb M}}
\def\ZZ{{\mathbb Z}}

\def\A{{\mathcal A}}
\def\B{{\mathcal B}}
\def\C{{\mathcal C}} 
\def\DHR{\C^{\mathrm{DHR}}}
\def\D{{\mathcal D}}

\def\H{{\mathcal H}}

\def\sig{\sigma}
\def\eps{\varepsilon} 
\def\inv{^{-1}} 
\def\LRA{\Leftrightarrow}

\def\wh{\widehat}
\def\ol{\overline}

\def\loc{_{\mathrm{loc}}}
\def\can{_{\mathrm{can}}}
\def\opp{^{\mathrm{opp}}}
\def\wick#1{{}:\!#1\!:{}}
\def\Psit{\Psi_{\tau\otimes\tau}}
\def\Psis{\Psi_{\sig\otimes\sig}}
\def\mm{\mathbf{m}}

\newcommand{\be}{\begin{equation}} 
\newcommand{\ee}{\end{equation}}
\newcommand{\ba}{\begin{array}} 
\newcommand{\ea}{\end{array}}
\newcommand{\bea}{\begin{eqnarray}} 
\newcommand{\eea}{\end{eqnarray}}

\newcommand{\eref}[1]{Eq.~(\ref{#1})}
\newcommand{\sref}[1]{Sect.~\ref{#1}}
\newcommand{\dref}[1]{Def.~\ref{#1}}
\newcommand{\cref}[1]{Cor.~\ref{#1}}
\newcommand{\lref}[1]{Lemma~\ref{#1}}
\newcommand{\rref}[1]{Remark~\ref{#1}}
\newcommand{\tref}[1]{Thm.~\ref{#1}}

\newcommand{\pref}[1]{Prop.~\ref{#1}}

\begin{document}

\title{\huge Phase boundaries in algebraic conformal QFT}
 
\author{{\sc Marcel Bischoff}\\ \scriptsize Institut f\"ur
  Theoretische Physik, Universit\"at G\"ottingen, 
\\[-1.5mm] \scriptsize Friedrich-Hund-Platz 1, D-37077 G\"ottingen,
Germany \\ \scriptsize Current address: Vanderbilt University, Department
of Mathematics \\[-1.5mm] \scriptsize 1326 Stevenson Center, Nashville, TN 37240,
USA \\[-1.5mm] \scriptsize {\tt marcel.bischoff@vanderbilt.edu}
\\[2mm]
{\sc Yasuyuki Kawahigashi} \\ \scriptsize Department of Mathematical Sciences, The
University of Tokyo, \\[-1.5mm] \scriptsize Komaba, Tokyo 153-8914, Japan; \\ \scriptsize Kavli IPMU
(WPI), The University of Tokyo \\[-1.5mm]
\scriptsize 5-1-5 Kashiwanoha, Kashiwa, 277-8583, Japan \\[-1.5mm] \scriptsize {\tt yasuyuki@ms.u-tokyo.ac.jp}
\\[2mm]
{\sc Roberto Longo} \\
\scriptsize Dipartimento di Matematica,
Universit\`a di Roma ``Tor Vergata'',\\[-1.5mm]
\scriptsize Via della Ricerca Scientifica, 1, I-00133 Roma, Italy \\[-1.5mm] \scriptsize {\tt longo@mat.uniroma2.it}
\\[2mm]
{\sc Karl-Henning Rehren} \\
\scriptsize Institut f\"ur Theoretische Physik, Universit\"at G\"ottingen,
\\[-1.5mm] \scriptsize Friedrich-Hund-Platz 1, D-37077 G\"ottingen, Germany
\\[-1.5mm] \scriptsize {\tt rehren@theorie.physik.uni-goettingen.de}}

\maketitle

\begin{center} \sl Dedicated to Detlev Buchholz on the occasion of his
  70th birthday.
\end{center}

\begin{abstract}
We study the structure of local algebras in
  relativistic conformal quantum field theory with phase boundaries. 
  Phase boundaries are instances of a more general 
  notion of boundaries that give rise to a variety of algebraic 
  structures. These can be formulated in a common framework
  originating in Algebraic QFT, with the principle of Einstein
  Causality playing a prominent role. We classify the  
  phase boundary conditions by the centre of a certain universal
  construction, which produces a reducible representation in which all
  possible boundary conditions are realized. For a large
  class of models, the classification reproduces results obtained in a
  different approach by Fuchs et al.\ before.  
\end{abstract}

\noindent
{\footnotesize Supported in part by the ERC Advanced Grant 669240
  QUEST ``Quantum Algebraic Structures and Models'', PRIN-MIUR,
  GNAMPA-INdAM and Alexander von Humboldt Foundation (RL). Supported
  by the Grants-in-Aid for Scientific Research, JSPS (YK). Supported
  by the German Research Foundation (Deutsche Forschungsgemeinschaft
  (DFG)) through the Institutional Strategy of the University of
  G\"ottingen (MB, KHR). 
  The hospitality and support of the Erwin-Schr\"odinger International
  Institute for Mathematical Physics, Vienna, is gratefully acknowledged. }

\section{Introduction}
\setcounter{equation}{0}

Boundaries and defects are ubiquitous phenomena in condensed matter
systems and have been extensively 
studies in the setting of Euclidean quantum field theory. In contrast, 
boundaries of relativistic quantum systems are less
studied, especially the impact of the principle of Causality that must
also be valid across the boundary. In view of the close 
(Osterwalder-Schrader) relation between Euclidean continuum field
theory and relativistic quantum field theory, the two issues are
expected to be closely connected, but the precise statement is not yet
clear. 

It is the aim of this work to develop a setup for the description 
of boundaries of relativistic quantum systems in the kinematically 
very simple case of two-dimensional conformal quantum field
theory. Indeed, the expected relation with the Euclidean setting shows
up in the relevant classification results. 

To fix the ideas, let us introduce our terminology. A boundary is a
timelike plane of co-dimension 1 in Minkowski spacetime, such as the 
plane $x=0$. In 2D, this is just the time axis. While a given
boundary itself necessarily breaks Poincar\'e invariance, in a 
relativistic description it can be moved to any other timelike plane. 

``The physics'' (i.e, the field content and the algebraic relations) 
may or may not be the same on both sides of the boundary. The extreme 
case is a ``system in a (one-sided) box'' with ``no physics'' on the 
other side of the boundary. We shall refer to such a boundary as a 
``hard boundary''. The opposite extreme has ``the same physics'' on 
both sides, with some discontinuity of the fields at the boundary. 
Between these extremes, there are many intermediate possiblities,
among them the ``phase boundaries'' (also called ``transmissive'' or
``transparent'' boundaries \cite{FFRS07}) which share the same
stress-energy tensor (and possibly other distinguished chiral fields)
on both sides, while the bulk field content may change. 

A ``boundary condition'' is, heuristically, a set of relations between the 
fields on both sides of the boundary. We shall give a technical
definition in \dref{d:bc}, and show in \sref{s:bcond} how it
accomplishes this heuristic idea.

The term ``defect'' is not used uniformly in the literature. 

In \cite{FRS03,FFRS04,TFT1,TFT2}, a defect is a transmissive boundary
condition between isomorphic or different two-dimensional (Euclidean)
CFTs on both sides. In \cite{FFRS07}, a conformal defect is a
boundary with energy conservation, including the reflective (hard
boundary) and the transmissive (phase boundary) case. 

An interesting independent approach to defects and boundaries in the
setting of nets of von Neumann algebras recently appeared in
\cite{BDH}. Here, defects are certain chiral nets of local algebras 
on $S^1$ split into two half-circles, that may involve degrees of
freedom that are attached to the boundary point, and are not required to be
covariant under translations away from the boundary. In fact, there is
no covariance assumed at all. For a more detailed comparison with our
\dref{d:bc}, see \rref{r:BDH} below. 

Because (in our setting) the two local theories at a phase boundary
share the same stress-energy tensor, they share the same unitary
representation of the conformal group on the Hilbert space. So the
field operators of the theory on one side of the phase boundary can be
translated to the other side by the translation unitaries -- where
they will in general fail to coincide with the fields of the theory on
the other side. It is important to notice that the Hilbert space
necessarily supports both sets of fields everywhere in Minkowski
spacetime, but on either side of the boundary only one of them
represents physical observables. The algebra generated by both sets of
fields will in general fail to be local.  

The first aim of this paper is to develop the algebraic framework to
describe quantum field theories with phase boundaries that are 
compatible with the Einstein Principle of Causality. We shall address 
this issue in the case of 2D conformal QFT, where the available mathematical
tools are most powerful, and where the most interesting classification
results can be established. 

The standard way to approach boundary problems is to {\em impose} 
boundary conditions, and study their implications. As it turns out in
2D CFT, one may indeed impose a continuum of boundary conditions on
the {\em chiral} fields. E.g., energy and / or momentum may be
required to be conserved. (A system in an energetically isolated box
has a hard boundary where energy is conserved, but momentum is not.) 

However, once a choice of the chiral boundary conditions is
made, one can no longer freely impose boundary conditions for the 
non-chiral bulk fields. Instead, these turn out to be ``quantized'' 
in a highly nontrivial way which can be traced back to algebraic
consistency conditions due to Einstein causality. Notice that this
is a feature that would be absent in classical field theory. It is
the main purpose of this paper to quantitatively describe the
quantization of boundary conditions for phase boundaries. The answer
will be given in terms of a generalized Verlinde formula pertaining to
the bimodule category of a modular tensor category (\sref{s:comp}). 
The case of hard boundaries has been treated earlier \cite{CKL13,LR04,LR09}. 

Our work is strongly influenced by, but at the same time largely
complementary to the work by Fuchs, Fr\"ohlich, Kong, Runkel,
Schweigert, notably \cite{TFT1,TFT2,FRS03,FFRS04,FFRS06,FFRS07,KR}. The latter 
line of research is situated in Euclidean quantum field theory, where
chiral fields arise as (anti-)~holomorphic fields defined on Riemann
surfaces. Lacking a ``Wick rotation'' applicable to QFT with
boundaries, it is not a priori clear that the Euclidean and the
relativistic approach should give the same classifications.  
Yet, in both cases, the mathematical analysis is based on the modular
tensor category of the chiral algebra. The properties of such
categories discovered along the way and giving rise to many highly
non-trivial and exciting mathematical structures, are explored, e.g.,
in \cite{FFRS06,KR,DKR}. They give rise to a ``topological quantum
field theory'' \cite{TFT1,TFT2} for partition functions and correlation
functions of 2D Euclidean CFT in terms of topological amplitudes
assigned to three-dimensional volumes bounded by two Riemann surfaces
that carry the CFT degrees of freedom.

The structures discovered in \cite{FFRS06} in the context of more
general ribbon categories, become instrumental for many of our
constructions and classifications where the underlying modular C*
tensor category is the DHR category of chiral superselection sectors
\cite{DHR,FRS,KLM}. On the other hand, our approach is complementary,
because it focusses on the physical interpretation in terms of local
algebras of observables, rather than correlation functions and
conformal blocks. We shall explain that Frobenius algebras in the
category describe extensions of the local von Neumann algebras of
observables, by controlling the algebraic relations that additional
fields on an extended Hilbert space must satisfy, consistent with
covariance and locality. Operations with Frobenius algebras (braided
products and decompositions, the ``centre'' and the ``full centre''
\cite{FFRS06,KR}) correspond to surprisingly elementary algebraic
operations like quotients of free products, central decompositions and
relative commutants of (nets of) local algebras \sref{s:Ext}. 

The common feature is the association of modules and bimodules with
boundary conditions. Because the relevant computation of the central
projections, that completes the classification of boundary conditions
for phase boundaries, is quite hidden in \cite{TFT1,TFT2,FFRS06,KR}, we
sketch here a new proof of \tref{t:center} that benefits
from some simplifications occurring in C* tensor categories,  
as opposed to the more general situation of tensor categories covered 
in the cited work. 

Namely, in our application to relativistic QFT, the relevant category
can be concretely realized as a subcategory of $\End_0(N)$ (the finite-index 
endomorphisms of a type $I\!I\!I$  von Neumann algebra $N$), and inherits the 
C* structure and notion of positivity from the latter. In compliance 
with earlier work on the classification of subfactors $N\subset M$ in
terms of Frobenius algebras in $\End_0(N)$ \cite{L94}, we shall use 
the term ``Q-system'' for a Frobenius algebra. In all that follows,
the important thing to be kept in mind is the one-to-one
correspondence between extensions of a local QFT and Q-systems in its
DHR category. 

The article is organized as follows. In \sref{s:Heur}, we present some
heuristics in a field theoretic language. In particular, we recall the
relativistic version of the critical Ising model. This model 
involves a pair of fields, called the ``order parameter'' $\sigma$ and 
the ``disorder parameter'' $\mu$, which each satisfy local commutation
relations with itself, but which satisfy certain non-local commutation 
relations with each other \cite{ST78}. They can therefore not arise
simultaneously in a QFT that satisfies the Principle of Causality. 
However, the commutation relations are such that $\sigma(x)$ commutes
with $\mu(y)$ if $x$ is in the left spacelike complement of $y$
(``left locality'').  
This situation is prototypical for a relativistic boundary condition:
if $\sigma$ belongs to the field content to the left of the boundary,
and $\mu$ to the right, then Causality is satisfied. Since $\sigma$
and $\mu$ are algebraically isomorphic, one has isomorphic field
algebras (``the same physics'') on both sides of the boundary, but a
discontinuity at the boundary. 

The bottom-line of these considerations is that one has to study
simultaneous extensions of a pair of local quantum field theories. 
Since every extension of a given field algebra yields a representation of
the latter by restriction, the issues of interest are of a
representation-theoretic nature. The best framework to address them is
therefore Algebraic Quantum Field Theory, which captures the
positive-energy representation theory in terms of the DHR
category of superselection sectors. This is a braided C* tensor 
category which we shall briefly review in \sref{s:SSS}. At this point, 
all the powerful categorical machinery becomes available. Ultimately,
we will be able to describe the two-dimensional boundary theory in terms of 
the DHR category of the underlying chiral algebra. We
  conclude the section with the precise definition of ``boundary
  conditions'' in the framework of Algebraic QFT.

In \sref{s:Ext}, we prepare the ground for the construction
  and classification of boundary conditions by
introducing the basic operations with possibly nonlocal 
extensions of local quantum field theories: The passage to maximal
local intermediate extensions, and 
the ``braided products'' of extensions which amounts to the merging 
of the charged fields of two extensions into a single extension. We 
present these operations as operations on local nets of von Neumann 
algebras in the AQFT setting, and relate them to the corresponding 
operations with Frobenius algebras (Q-systems).

The braided product can be understood as a quotient of a free product of 
two extensions of the same underlying chiral algebra by the obvious 
relations due to the identification of the common chiral algebra, 
and by a single additional relation imposing left or right locality. 
It therefore necessarily violates ``one half'' of locality. New local 
models can be obtained by passing to one of the maximal local
intermediate extensions.

In particular, a simple algebraic understanding of the previous
``$\alpha$-induction construction'' of \cite{R00} emerges: Its local
algebras are relative commutants of algebras of a nonlocal braided
product, associated with wedge regions. This puts the
$\alpha$-induction construction on the same footing with the
construction of hard boundary theories in terms of relative
commutants \cite{LR04}.  

In \sref{s:Bound}, we return to the issue of boundary conditions at
phase boundaries. Recall that the bulk algebras on both sides 
of the boundary are two local extensions of the common chiral subtheory, 
defined on the same Hilbert space, with the left extension being 
left-local w.r.t.\ the right extension. But these are precisely the 
relations defining the braided product of two local extensions
(obtained by the appropriate product of Q-systems) as a quotient of 
the free product, and there is no need to descend to a maximal local
intermediate extension. The main structural result in this
  section is the observation (\pref{p:universal}) that the braided
  product algebra is a ``universal construction'', that contains all
  irreducible boundary conditions.

It turns out that the braided product of two local extensions in
general has a centre, and its irreducible representations are in 
one-to-one correspondence with its minimal central projections
(\cref{c:bc=mp}).  
Thus, in order to classify boundary conditions, the task is to 
identify the minimal central projections in a braided product 
of commutative Q-systems. In \sref{s:bcond}, we
reformulate this problem in terms of a commutative product on a
finite-dimensional intertwiner space.

That this task can be completely accomplished in a
  special case of prominent interest, is another main result of the present
  work: in \tref{t:center}, we classify the irreducible 
  boundary conditions in the case when the Q-systems of the two local
  nets arise by the $\alpha$-induction construction
  from two chiral Q-systems $A$, $B$ in a modular tensor
  category. These are the maximal two-dimensional extensions of
  completely rational chiral quantum field theories.      
Here is in fact the only point where modularity of the category is 
required. The finding is that the minimal projections are in 
one-to-one correspondence with $A$-$B$-bimodules between the pair of 
chiral Q-systems. The central projections implement certain relations 
between the fields on both sides of the boundary (e.g., they may coincide, 
flip sign, or stand at certain angles to each other). These are the 
``quantized boundary conditions'' mentioned earlier. The precise
values of the angles are given by matrix elements of generalized
Verlinde matrices that diagonalize the bimodule fusion rules. 

This classification in terms of chiral bimodules was in fact
previously discovered in the framework of two-dimensional Euclidean
CFT \cite{TFT1,TFT2,FFRS04,FFRS06,FFRS07,KR}. Our streamlined proof 
(\cite[Thm.\ 4.44]{BKLR}) takes advantage of the simplifications
occurring in C* tensor categories, by using a positivity argument in
the decisive step establishing orthogonality of the projections,
cf.\ the paragraph after \tref{t:center}).

Along the way, we shall completely work out the example of the
two-dimensional Ising model (\sref{E4}, \sref{E5}, \sref{E6}), for
which the algebraic computations are sufficiently simple to be done
explicitly.

\section{Heuristic considerations}
\setcounter{equation}{0} \label{s:Heur}

We rephrase in a Lorentzian setting the characterization of
\cite{FFRS07}, which is placed in a Euclidean setting. Spacetime is
two-dimensional Minkowski space with coordinates $(t,x)$. Without loss
of generality, we place the boundary at $x=0$, so the boundary is the
time axis in $\MM_2$. 

\subsection{Boundary conditions for chiral fields}
We denote the subspace $x>0$ by $\MM_R$, and the
subspace $x<0$ by $\MM_L$. On both sides of the boundary, we have 
left- and right-moving chiral fields, for which we reserve the labels 
$+$ and $-$\footnote{corresponding to holomorphic and anti-holomorphic
  fields in the Euclidean framework. The renaming of ``$M_+$'' in \cite{LR04}
  into $\MM_R$ in this paper is made in order to prevent confusion.}.

Let us accordingly denote the chiral components of the stress-energy tensor 
on either side of the boundary by $T^L_+(t+x)$ , $T^L_-(t-x)$ ,
$T^R_+(t+x)$ , $T^R_-(t-x)$ (where the former two are defined at
$x<0$, and the latter two at $x>0$). Then the energy conservation at
the boundary ($T^L_{01}(t,0)=T^R_{01}(t,0)$) reads\footnote{This condition is 
called ``conformal defect'' in \cite{FFRS07}, whereas the more special
case \eref{trans} is called ``topological''. In \cite{FFRS04},
\eref{trans} is referred to as ``conformal''.}  
\be\label{econs}
T^L_+(t)+T^R_-(t)= T^R_+(t)+T^L_-(t).\ee
There are several types of solutions to this condition: the {\bf reflective}
solution  
$$T^L_+(t)=T^L_-(t),\qquad T^R_+(t)=T^R_-(t),$$
the {\bf transmissive} solution
\be\label{trans}
T^L_+(t)=T^R_+(t),\qquad T^L_-(t)=T^R_-(t),
\ee
or combinations of both, like 
$$T^L_+=T_1+T_2, \quad T^R_-=T_3+T_4,\qquad T^R_+=T_1+T_3, \quad
T^L_-=T_2+T_4,$$
where $T_k$ are independent chiral stress-energy tensors, and $T_1$
and $T_4$ are transmitted while $T_2$ and $T_3$ are reflected. 
There are more solutions which are of neither type, eg, if $j_1$ and 
$j_2$ are two independent chiral currents, then  
$$T^L_+=j_1^2,\quad T^R_-=j_2^2,\qquad T^R_+=(\cos\alpha
j_1+\sin\alpha j_2)^2,\quad T^L_-=(-\sin\alpha
j_1+\cos\alpha j_2)^2$$
for any angle $\alpha$ also solve the condition \eref{econs}. In this case, the
boundary effect is a gauge transformation of the $O(2)$-symmetric
doublet $(j_1,j_2)$.  

Since causality should also hold in the presence of a boundary, one
obtains conditions on the local commutativity among the four
stress-tensors involved. (The argument is standard, 
exploiting the fact that a chiral field, say $\phi_+(u)$, is localized 
at any point $(t,x)$ such that $t+x=u$, with the restriction that, 
say for $T^R_\pm$, $x$ must be positive. Whenever one can find pairs 
of two such points which are spacelike separated, then the two field 
operators must commute.) First considering points at equal time in 
$\MM_R$, one finds that both $T^R_\pm$ must be local fields (i.e.,
commute with themselves except at coinciding points), and that 
$T^R_+$ must be forward-local w.r.t.\ $T^R_-$ (i.e.,
$[T^R_+(u),T^R_-(v)]=0$ whenever $u>v=t-x$). Similarly, $T^L_\pm$ must be
local, and $T^L_-$ must be forward-local w.r.t.\ $T^L_+$. Considering
points at equal time on opposite sides of the boundary, one finds that
$T^R_+$ must commute with $T^L_-$, and $T^L_+$ must commute with
$T^R_-$ (that is: at all points, also coinciding ones). 

But requiring also M\"obius covariance of the chiral fields,
forward locality implies backward locality because chiral fields on $\RR$
extend to the compactification $S^1$, so the two mutually commuting 
fields $T^R_+$, $T^L_-$ must in fact be relatively local w.r.t.\ to
the two mutually commuting fields $T^L_+$, $T^R_-$ (that is, the
commutator is supported at equal points). Thus, all four fields are
mutually local w.r.t.\ each other, and $T^L_+$ commutes with $T^R_-$,
and $T^R_+$ commutes with $T^L_-$. These conditions are met in
all the examples above.   

In the reflective case, it follows that the two sides completely
decouple, and the theory may be regarded as a tensor product of two
hard boundary theories on either side, living on independent Hilbert
spaces. 

On the other hand, the transmissive solution \eref{trans} may also be 
characterized by the second condition supplementing \eref{econs}
\be
\label{pcons}
T^L_+(t)-T^R_-(t)= T^R_+(t)-T^L_-(t).
\ee
This equation just expresses the conservation of momentum at the
boundary. (Clearly, momentum is not conserved at a hard boundary.)
While we always want to impose energy conservation by default,
momentum conservation is equivalent to the transmissivity condition
for the stress-energy tensors. Notice that the commutativity between 
$T^L_+$ and $T^R_-$ then implies that $T^L_+=T^R_+$ commutes with 
$T^L_-=T^R_-$.

In general, there may be other chiral fields. In fact, every 
two-dimensional conserved rank-$n$ symmetric traceless tensor current
splits into a pair of two chiral fields of scaling dimension $n$. 
Thus one can imagine that by imposing or not imposing conservation of
the associated charge at the boundary, one obtains a vast multitude of
possibilities, among which hard boundaries (reflective for all chiral
fields) and phase boundaries (transmissive for all chiral fields) are
only extremal cases.

In general, the boundary will be transmissive for a subset of
chiral fields if it preserves the corresponding symmetries, so that the 
two field theories on both sides of the boundary share the transmitted 
chiral fields.

\subsection{Bulk fields}
\label{s:bulk}

Apart from the transmitted (hence common) chiral fields, the quantum 
field theories on each side may have more chiral fields that are not 
transmitted, and each side may have further
local fields that are not chiral. Thus both sides are extensions of
quantum field theories of the form $\A_+\otimes \A_-$, standing for
the algebras of the transmitted left- and right-moving chiral fields
on the respective lightrays $\RR$.

Extensions of local field theories have been studied in the
operator-algebraic setting, starting with \cite{DHR}, and further 
developped for chiral conformal quantum field theories in \cite{LR95} 
generalizing and conceptualizing the notion of ``conformal embeddings'' 
\cite{SW}. The crucial assumption is that the relative commutant of
the subtheory in the extension is trivial; this ensures that the
subtheory contains the total stress-energy tensor of the full 
theory, since otherwise the relative commutant is a QFT of its own 
(the coset theory \cite{GKO} in the chiral case) which commutes with 
the given subtheory. This assumption is certainly not true in the case 
of a hard boundary, where the transmitted part of the stress-energy 
tensor is zero. On the other hand, if $\A_+\otimes \A_-$ is
nontrivial, there may exist different extensions with trivial relative
commutant which qualify as different ``phases'' describing the
possible local quantum field theories on either side of a transmissive 
boundary. For this reason, we also refer to transmissive boundaries as
{\bf phase boundaries}. 

In particular, the theory of local extensions \cite{LR95} is
applicable in the case of phase boundaries, including all energy and 
momentum conserving boundaries.

The transmitted left- and right-moving chiral fields $\A_+$ and $\A_-$
may be specified independently. Most of the literature refers to 
the case when they are in fact isomorphic. This may be justified by
the option to have several boundaries, including hard ones: As we have 
seen, a hard boundary on one side forces the right-moving and left-moving 
fields to coincide in the bulk between this boundary and a
transmissive boundary on the other side. 

Only for special issues: the canonical local extension
(\sref{s:canonical}), the $\alpha$-induction construction
(\sref{s:alpha}), and the complete classification of irreducible
boundary conditions in terms of bimodules (\sref{s:classi}), we shall limit  
our treatment to the case with isomorphic left- and right-moving 
chiral fields, i.e., we shall regard the bulk CFT on both sides of 
the boundary as (different) local extensions of the same chiral 
subtheory of the form $\A\otimes \A$ with a fixed chiral theory $\A$. 
In this case, it is well known that there is always a ``canonical'' 
local extension of $\A\otimes\A$ (\pref{p:canonical}), 
while more general extensions can be obtained by the
$\alpha$-induction construction \cite{R00}. In fact, every extension 
with trivial relative commutant is intermediate between the trivial 
extension ($\B_{\rm 2D}=\A\otimes\A$) and one of the $\alpha$-induction 
extensions \cite{LR09}.

\subsection{Left locality} \label{s:ll}
In the presence of a transmissive boundary, denote the observable fields 
to the left and to the right of the boundary by $\Psi^L$ and $\Psi^R$. 
Both field algebras contain the common chiral stress-energy tensor $T_\pm$.  

Since the stress-energy tensor provides the densitites for the generators of 
space and time translations, the fields $\Psi^L$ and $\Psi^R$, defined 
on either side of the boundary, in fact extend to the full Minkowski 
spacetime by means of translations. However, on the ``wrong side'', they 
are not necessarily observables. 

The Principle of Causality requires that observables at spacelike 
separation commute. Hence $\Psi^L(x)$ must commute with $\Psi^R(y)$ 
whenever $x$ is to left of the boundary, $y$ is to the right of 
the boundary, and $(x-y)^2<0$. Using translations, the same must be 
true whenever $x$ is to the spacelike left of $y$. This motivates the 

\begin{definition} A quantum field theory with fields $\Psi^L$ is 
called {\bf left-local} w.r.t.\ a QFT with fields $\Psi^R$ defined on
the same Hilbert space, if 
$$[\Psi^L(x),\Psi^R(y)]=0$$
whenever $x$ is to the spacelike left of $y$.  
\end{definition}

Then, the construction of a transmissive boundary between two given 
quantum field theories amounts to finding a representation of both 
algebras of fields $\Psi^L$ and $\Psi^R$ on the same Hilbert space 
such that $\Psi^L$ is left-local w.r.t.\ $\Psi^R$, and both
stress-energy tensors are represented by the same operators.

\subsection{Example 1: The Ising model}
\label{E1}

The prototype of a phase boundary may occur in the Ising model. We
briefly review the basic results for the relativistic Ising model, as
studied by \cite{ST78}. Recall 
that the local fields of the Ising model are the two chiral components
of stress-energy tensor with $c=\frac12$, a local field $\eps(t,x)$ with
scaling dimensions $(\frac12,\frac12)$, and another local field $\sig(t,x)$
with scaling dimensions $(\frac1{16},\frac1{16})$. The theory may be
extended by the chiral Fermi fields $\psi_+(t+x)$ and $\psi_-(t-x)$ on
a larger Hilbert space, such that 
\bea\label{algR}
\eps(t,x)=-i\psi_+(t+x)\psi_-(t-x).
\eea
(The imaginary factor is necessary to make $\eps$ hermitean when
$\psi_\pm$ are hermitean.) In particular, both $\psi_\pm$ are
relatively local w.r.t.\ to $\eps$. In contrast, they are not relatively
local w.r.t.\ to $\sigma(t,x)$: one has instead the commutation relations
at spacelike distance (displayed at equal times $t=0$)
\bea
\label{dualCR1}
\sig(0,x)\psi_+(y)=\pm \psi_+(y)\sig(0,x),\nonumber \\
\sig(0,x)\psi_-(-y)=\pm\psi_-(-y)\sig(0,x),
\eea 
with the sign given by $\pm =\mathrm{sign}(x-y)$. (Recall that by our
coordinate conventions, $\psi_\pm(\pm y)$ are both localized at the point 
$(0,y)$.) One may finally introduce the (short-distance regularized) operator
product 
\bea\label{dualF}
\mu(t,x)=\sqrt i\wick{\psi_+(t+x)\sig(t,x)} 
= \sqrt {-i}\wick{\psi_-(t-x)\sig(t,x)}.
\eea 
This field is known as the {\em dual field}. Indeed, the fields 
$T_\pm,-\eps,\mu$ have exactly the same correlation functions as the 
fields $T_\pm,\eps,\sigma$. (In the Euclidean version of the Ising
model, $\mu$ is known as the order parameter, indicating the phase 
transition at the critical temperature.) Conversely, 
\bea\label{dualF2}
\sig(t,x)=\sqrt i\wick{\psi_-(t-x)\mu(t,x)} 
= \sqrt {-i}\wick{\psi_+(t+x)\mu(t,x)}.
\eea 

Now, a quantum field theory with a phase boundary at $x=0$ is given as
follows. Its generating local fields on the right halfplane are
$T_\pm,\eps,\sigma$, and its local fields on the left halfplane are 
$T_\pm,\eps,\mu$. By the commutation relations \eref{dualCR1}, 
the field $\mu$ is left-local w.r.t.\ the field $\sig$. In particular,
\bea
\label{dualCR}
\mu(x)\sig(y)=\pm\sig(y)\mu(x) \qquad ((x-y)^2<0, x\in \MM_L, y\in \MM_R).
\eea 
Thus, the theory satisfies causality also across the phase boundary.

Because both $\sig$ and $\mu$ must be represented on the same Hilbert 
space, the common Hilbert space is the enlarged Hilbert space that 
accomodates also the chiral Fermi fields. (A detailed analysis of the 
structure of this Hilbert space has been presented in \cite{FR98}. As a 
representation of the two chiral Virasoro algebras, it has the form
\be\label{dualhilbert}
\H=(\H_0\oplus\H_{\frac12})\otimes(\H_0\oplus\H_{\frac12})\oplus 2
(\H_{\frac1{16}}\otimes\H_{\frac1{16}}),
\ee 
where the representation $\frac1{16}\otimes\frac1{16}$ occurs twice.)

Finally, by the equality of the correlation functions as stated above,
the local field algebras on both sides of the boundary are perfectly 
isomorphic, though different subalgebras of the nonlocal algebra
generated by the order and disorder fields \eref{dualF}--\eref{dualF2}.

\section{Theory of superselection sectors}
\label{s:SSS} \setcounter{equation}{0}

The classification 
of the possible boundary conditions is an issue of representation
theory. For this kind of questions, the setting of algebraic quantum
field theory is the most appropriate \cite{DHR}. Apart from the
description of representations in terms of endomorphisms,  
which allows to exploit the full power of the theory of
subfactors, it also has the benefit of sparing the necessity of 
short-distance regularizations of operator products, and other issues 
of technical nature. 

We shall give a brief review of the underlying DHR theory and the use
of Q-systems in order to characterize extensions of a given QFT. A
Q-system is essentially the same thing as a Frobenius algebra in a
tensor category, with the additional qualification that the tensor
category carries a natural C* structure, inherited from the local
von Neumann algebras. This feature simplifies some of the general
results about Frobenius algebras. For us, the important point is that
a Q-system allows to grasp all algebraic features of an extension
(commutation relations, operator product expansion) in terms of
finitely many data, as we will explain below. 

In the sequel of the paper (\sref{s:Ext}), we shall review various
operations with Q-systems, known from purely categorical approaches to
CFT, and establish their meaning in terms of elementary algebraic
operations. It then turns out
(\sref{s:Bound}) that boundaries are described by braided product Q-systems
which precisely ensure one-sided locality. 

\subsection{Algebraic QFT}

Algebraic quantum field theory starts from the net of local
algebras. This means the assignment $O\mapsto\A(O)$ where $O$ is a
spacetime region, and $\A(O)$ the algebra of observables localized in
$O$. The net is subject to the Haag-Kastler axioms, which just 
``transcribe'' the Wightman axioms without referring to individual 
fields as generators of the local algebras. Instead, the latter may 
be taken as von Neumann algebras (provided the causal complement of 
$O$ is nontrivial). In the case at hand, we assume two nets of
(left- and right-moving) chiral algebras, indexed by the open
intervals $I\subset\RR$ of the real line (= future-directed light
rays). A two-dimensional quantum field theory is then an extension 
$$\A_+(I)\otimes\A_-(J)\equiv \A_{\rm 2D}(O)\subset \B_{\rm 2D}(O),$$
where 
$$O=I\times J=\{(t,x):t+x\in I,t-x\in J\}$$ 
are the double cones in Minkowski spacetime $\MM_2$. 
Extensions are assumed to be covariant, {\em relatively local}, i.e., 
$\B_{\rm 2D}(O_1)$ commutes with $\A_{\rm 2D}(O_2)$ if $O_1$ and $O_2$ are spacelike
separated, and irreducible, i.e., the relative commutant $\A_{\rm 2D}(O)'\cap
\B_{\rm 2D}(O)$ is trivial. If $\B_{\rm 2D}$ is itself local, we call it a 
{\em local extension}. 

At some point, we shall be forced to admit finite-dimensional relative 
commutants. This obviously still excludes the trivial possibility of 
extensions via tensor products with independent theories. 
In particular, the ``coset stress-energy tensor'' \cite{GKO} is
trivial, because it is affiliated with the relative commutant, hence 
$\B_{\rm 2D}$ and $\A_{\rm 2D}$ share the same stress-energy tensor (i.e., a common 
Virasoro subnet), and the generators of the local diffeomorphisms 
of $\A_{\rm 2D}$ also generate the local diffeomorphisms of $\B_{\rm 2D}$. 

Every irreducible (positive-energy) representation of
$\B_{\rm 2D}$ restricts to a (positive-energy) representation of $\A_{\rm 2D}$. The
latter may be reducible. This explains why extensions are an issue of
representation theory. 

\subsection{Review of DHR theory for chiral conformal QFT}

By default, we shall assume Haag duality for the chiral observables. 
This means $\A(I)=\A(I')'$, where $I'$ is the complement in $\RR$ 
(= the union of two halfrays). (On the conformal completion $S^1$ of
$\RR$, Haag duality is automatic, whereas on $\RR$ it is a 
stronger property, equivalent to ``strong additivity''.)
This implies \cite{BMT} that all positive-energy representations are given 
by ``DHR endomorphisms'' $\rho$ of the net \cite{DHR}, namely
endomorphisms of $\A$ such that, up to unitary equivalence 
$$\pi\equiv\pi_0\circ\rho,$$ 
where $\pi_0$ is the vacuum representation. The DHR endomorphisms are
localized in some interval $I$ in the sense that $\rho(a)=a$ for all
$a\in \A(I')$. The localization interval may be varied arbitrarily,
namely for every other interval $\wh I$, one can chose unitary
``charge transporters'' such that $\wh\rho(\cdot)=
\Ad_{u}\rho(\cdot)=u\rho(\cdot)u^*$ is localized in $\wh I$.  

Intertwiners between localized endomorphisms $\rho$, $\sig$ are
operators $t$ satisfying $t\rho(a)=\sig(a)t$ for all $a\in\A$. We
write $t:\rho\to\sig$ or $t \in \Hom(\rho,\sigma)$. By Haag duality, every intertwiner
$t:\rho\to\sig$ between two localized endomorphisms is an element of
some local algebra, hence $t_1\rho_1(t_2)=\sig_1(t_2)t_1$ is an
intertwiner $\rho_1\rho_2\to\sig_1\sig_2$. 

The composition of DHR endomorphisms defines a ``fusion'' product of 
representations. The composition of localized endomorphisms is
commutative when the localization intervals do not overlap. In
general, it is commutative only up to unitary equivalence, implemented
by unitary statistics operators (``braiding'') 
$$\eps_{\rho,\sig}:\rho\sig\to\sig\rho.$$
In chiral theories, the braiding is defined \cite{FRS} by
$$\eps_{\rho,\sig}:= \sig(u_\rho^*)u_\sig^*u_\rho\rho(u_\sig)\quad 
:\rho\sig\to\rho\wh\sig\to\wh\rho\wh\sig=
\wh\sig\wh\rho\to\sig\wh\rho\to\sig\rho,$$
where $u_\rho$, $u_\sig$ are unitaries such that the auxiliary DHR
endomorphism $\wh\rho=\Ad_{u_\rho}\rho$ and 
$\wh\sig(\cdot) = \Ad_{u_\sig}\sig$ are localized in auxiliary
intervals $I_\rho$, $I_\sig$ such that $I_\sig < I_\rho$ (``$\sigma$
is localized in the past of $\rho$''). The braiding is independent
of the auxiliaries with the given specifications. In particular, one has 
\be\label{trivbraid}
\eps_{\rho,\sig}=\mathbf{1} \qquad\hbox{whenever $\sig$ is localized in the 
past of $\rho$ }
\ee
(because one may just take $u_\rho=u_\sig=\mathbf{1}$). Actually,
\eref{trivbraid} may be regarded as an ``initial condition'' which
completely fixes the braiding in the general case. The opposite braiding 
$\eps^-_{\rho,\sig}\equiv\eps_{\sig,\rho}^*$ would have resulted with
the opposite ordering of the auxiliary intervals. We also write $\eps^+\equiv\eps$. 

These structures and data turn the DHR representation theory of $\A$
into a braided C* tensor category $\C=\DHR(\A)$. This category is
simple ($\Hom(\id,\id)=\CC$), strict (the composition is associative),
and direct sums and subobjects corresponding to non-zero projections
$p\in\Hom(\rho,\rho)$ are defined.

We shall throughout assume complete rationality, 
which implies that $\A$ possesses only finitely many inequivalent 
positive-energy representations, and all of them have conjugates and finite
statistical dimension \cite{LX}. In this case, we call $\DHR(\A)$
``rational''. However, complete rationality of $\A$ becomes
truely essential only for the final classification of the boundary
conditions in \tref{t:center}, because it ensures modularity of $\DHR(\A)$. Otherwise,
one may as well choose $\C$ to be any rational subcategory of
$\DHR(\A)$.

\subsection{Example 2: The Virasoro net with $c=\frac12$ (= the
  ``chiral Ising model'')} \label{E2}

The chiral Ising model (=$\Vir(c=\frac12)$) has three sectors: $\pi_0$
(the vacuum representation), $\pi_\tau$ and $\pi_\sigma$ satisfying
the fusion rules $\tau\times\tau \simeq\id$,
$\tau\times\sig\simeq\sig\times\tau\simeq\sig$, $\sig\times\sig
\simeq\id\oplus\tau$. These are realized by localized endomorphisms $\id$,
$\tau$, $\sig$. By choosing $\tau$ in its unitary
equivalence class such that $\tau\sig=\sig$, it follows that
$\tau^2=\id$ and $\sig\tau=\Ad_u\sig$, where
$u\in\Hom(\sig,\sig\tau)=\Hom(\sig\tau,\sig)$ is a unitary with
$u^2=\mathbf{1}$. $u$ is not a multiple of $\mathbf{1}$ because otherwise
$\sig\tau=\tau$ would imply $\tau=\id$.
One may choose a pair of
intertwiners $r\in\Hom(\id,\sig^2)$, $t\in\Hom(\tau,\sig^2)$
satisfying the Cuntz relations
$r^*r=\mathbf{1}$, $t^*t=\mathbf{1}$, $r^*t=0$, $rr^*+tt^*=\mathbf{1}$. Since
$u\in\Hom(\sig,\sig\tau)\subset \Hom(\sig^2,\sig^2)$ which is spanned by
$rr^*$ and $tt^*$, it follows that $u=rr^*-tt^*$ (fixing its sign). 

Now, the action of $\tau$ and $\sig$ on $r$ and $t$ can be computed as
follows. $\tau(r)\in\Hom(\tau,\tau\sig^2)=\Hom(\tau,\sig^2)$ is a
multiple of $t$, hence $\tau(r)=t$ (fixing a relative complex phase),
and $\tau(t)=r$. Likewise,
$\sig(r)\in\Hom(\sig,\sig^3)=(rr^*+tt^*)\Hom(\sig,\sig^3)=r\cdot
r^*\Hom(\sig,\sig^3)+t\cdot t^*\Hom(\sig,\sig^3) =r\cdot\Hom(\sig,\sig)+t\cdot \Hom(\sig,\tau\sig) = r\cdot \CC+t\cdot\CC$,
and similarly
$\sig(t)\in\Hom(\sig\tau,\sig^3)=ru\cdot\CC+tu\cdot\CC$. On the other 
hand, $\sig$ must preserve the Cuntz relations of the
isometries, and intertwining relations like $\sig^2(r)t=t\tau(r)$ etc.\ must hold. 
In other words, the defining relations of a tensor category constitute
algebraic relations for the unknown coefficients, that can be solved.   
(This way of exhibiting a C* tensor category is known as the
``Cuntz algebra construction'' \cite{I,EG}.) The same strategy also allows
to compute the braiding operators. The result for the Ising model is
\bea \tau(r)=t,\quad  \tau(t)=r,\quad  \tau(u)=-u,\hskip37mm\notag
\\ \sig(r)=(r+t)/\sqrt2,\quad  \sig(t)=(r-t)u/\sqrt2,\quad 
\sig(u)=rt^*+tr^*,\hskip17mm 
\\ \eps_{\tau,\tau}=-1,\quad\eps_{\tau,\sig}=\eps_{\sig,\tau}=-iu,
\quad\eps_{\sig,\sig}=\kappa_\sig^{-1}(rr^*+itt^*)\qquad 
(\kappa_\sig=e^{2\pi ih_\sig}=e^{2\pi i/16}).\notag\eea

There exist in fact eight solutions for a braided tensor category with
the given fusion rules, describing the DHR categories
of other models whose sectors share the same fusion rules, e.g., 
the $su(2)$ current algebra at level 2. The correct solution is 
distinguished by the statistics parameter defined as the complex phase
of $r^*\sig(\eps_{\sig,\sig})r \in\CC\cdot\mathbf{1}$ \cite{FRS}
whose value is known (by the Spin-Statistics theorem \cite{GL96}) from
the scaling dimension of the sector $\sig$.

\subsection{Two dimensions}

In two dimensions, the analysis is essentially the same, replacing
intervals by double cones, and taking $O'$ to be the causal complement
in $\MM_2$ (= the union of two wedge regions). The braiding is defined
such that 
\be\label{trivbraid2}
\eps_{\rho,\sig}=\mathbf{1} \qquad\hbox{whenever $\sig$ is localized in the 
spacelike left of $\rho$.}
\ee
In particular, the irreducible localized endomorphisms of 
$\A_{\rm 2D}=\A_+\otimes\A_-$ are tensor products $\rho=\rho^+\otimes\rho^-$, 
and, in view of \eref{trivbraid}, the convention \eref{trivbraid2}
means that the two-dimensional braiding is 
$$\eps_{\rho,\sig}= \varepsilon_{\rho^+,\sigma^+}^+ \otimes
\varepsilon_{\rho^-,\sigma^-}^- \equiv\eps_{\rho^+,\sig^+}\otimes\eps_{\sig^-,\rho^-}^*.$$

The DHR category of $\A_{\rm 2D}=\A_+\otimes\A_-$ is the completion
under direct sums $\DHR(\A_{\rm 2D}) = \DHR(\A_+)\boxtimes\DHR(\A_-)\opp$
of the tensor product of categories
$\DHR(\A_+)\otimes\DHR(\A_-)\opp$. Here, we write $\C\opp$ for the tensor category equipped with the opposite braiding
$\eps^-$. 

\subsection{Extensions}
\label{s:extensions}

An extension of a local QFT $\A$ is a quantum field
theory $\B$ together with an embedding $\iota:\A\to\B$ such that
$\iota(\A(O))\subset \B(O)$ for every region $O$, and $\B$ is
relatively local w.r.t.\ $\A$ (but not necessarily itself local).

In \sref{s:Ext}, we shall describe finite-index extensions in terms of
the DHR category $\DHR(\A)$ (making essential use of the fact that the vacuum
representation of $\B$ restricts to a DHR representation $\theta$ of $\A$). 
Heuristically, the condition of finite index ensures that $\A$ and
$\B$ share the same stress-energy tensor.
If $\A$ is rational, all irreducible extensions have finite
index \cite{LR95}. Otherwise, the same is true if $\theta$ belongs to
some rational full subcategory of $\DHR(\A)$. 

We shall introduce in \sref{s:charged}
``charged field operators'' $\Psi_\rho\in\B$ that 
intertwine a DHR representation $\rho\prec \theta$ of $\A$ with the vacuum
representation, and together with the subalgebra $\A$ generate
$\B$. These are the AQFT counterparts of sector-generating primary
fields in CFT.

\subsection{Phase boundaries}
\label{s:phaseb}

A phase boundary is given by specifying two {\em local} extensions
$\B^L_{\rm 2D}$ and
$\B^R_{\rm 2D}$ of a common (completely rational) subnet
$\A_{\rm 2D}=\A_+\otimes\A_-$ of chiral observables,
covariantly represented on the same Hilbert space, such that
$$\B_{\rm 2D}(O)=\left\{\ba{l}\B_{\rm 2D}^L(O)\\[1.5mm] \B_{\rm 2D}^R(O)\ea\right.\quad\hbox{if $O$
  lies to the} \left\{\ba{l}\hbox{left} \\ \hbox{right}\ea\right. 
\hbox{of the boundary}.$$
The choice of $\A_+$ and $\A_-$ is part of the specification of the
model; as motivated by \eref{trans}, $\A_\pm$ are assumed to contain
the generators of the local diffeomorphisms, and hence the local
algebras of both nets $\B^L_{\rm 2D}$ and $\B^R_{\rm 2D}$ are in fact  
{\em defined} for every region $O\subset\MM_2$. They just do not
qualify as observables on the ``wrong'' side of the
boundary. Causality requires that the algebras  
$\B^L_{\rm 2D}(O_1)$ and $\B^R_{\rm 2D}(O_2)$ commute whenever $O_1\subset \MM^L$
and $O_2\subset \MM^R$ are spacelike separated. By diffeomorphism 
covariance, the same must be true whenever $O_1$ is to the spacelike 
left of $O_2$. 

Thus, the actual location of the boundary is perfectly arbitrary, 
and it may be any timelike curve. This feature is referred to as 
``topological'' in \cite{TFT1,TFT2}. 

As in \sref{s:ll}, we define ``left locality'':

\begin{definition}\label{leftlocal}
For two nets of local algebras $\B_i(O)$ defined on the same Hilbert
space, we say that $\B_1$ is {\bf left-local} w.r.t.\ $\B_2$, if $\B_1(O_1)$
commutes with $\B_2(O_2)$ whenever $O_1$ is in the left causal
complement of $O_2$. $\B_1$ is {\bf right-local} w.r.t.\ $\B_2$, iff 
$\B_2$ is left-local w.r.t.\ $\B_1$. The same definition makes sense
for a pair of chiral nets, replacing double cones $O$ by intervals
$I\subset\RR$, except that in this case the terms ``backward-local'' and
``forward-local'' may be more appropriate \cite{LR12}. 
\end{definition}

We shall treat chiral and two-dimensional phase boundaries at an
equal footing, and suppress the subscript $2D$ in the remainder of
this section.

Thus, at a phase boundary, $\B^L$ has to be left-local w.r.t.\
$\B^R$. Given a phase boundary, one may consider 
the local algebras 
\be
\label{eq:add}
\D(O):=\B^L(O)\vee\B^R(O).
\ee
$\D$ is another extension of $\A$, but $\D$ will in general be
nonlocal, because $\B^L$ may (and will) not be also right-local w.r.t.\
$\B^R$, cf.\ Example 1, \sref{E1}. A boundary {\em condition} then
specifies the precise ``relative algebraic position'' of the local
subtheories $\B^L$ and $\B^R$ within $\D$. 

\begin{definition} \label{d:bc}
A {\bf defect} between nets $\B^L\supset \A$ and $\B^R \supset \A$
with respect to $\A$ is implemented by a relatively local 
extension $\D \supset \A$ of finite index, such that $\B^{L/R}$
are intermediate extensions and $\D$ is left-local
  w.r.t.\ $\B^R$ and right-local w.r.t.\ $\B^L$. In particular, 
$\B^L$ is left-local with respect to $\B^R$. 
We call it a \textbf{boundary condition} if \eref{eq:add} holds, 
\textbf{simple} or \textbf{factorial} if $\D(O)$ are factors, and 
\textbf{irreducible} if $\A\subset \D$ is an irreducible extension, 
i.e., $\A(O)'\cap \D(O)=\CC\cdot\mathbf{1}$.
\end{definition}
If $\A\subset\D$ is a defect between $\A\subset\B^L$ and
$\A\subset\B^R$, then the restriction to $\A\subset\B^L\vee\B^R$ is a
boundary condition. For the justification of the terminology ``simple $\equiv$ factorial'',
see \sref{s:decoQ}.

\begin{remark}
The above definition applies as well for boundary conditions at 
one-dimensional phase boundaries in two-dimensional Minkowski 
spacetime (our actual issue of interest), as for boundary conditions
between chiral nets where the boundary is just a point. 
\end{remark}

\begin{remark} \label{r:BDH}
As we shall see in \sref{s:bcond}, the condition \eref{eq:add} ensures
that irreducible boundary conditions can be formulated as algebraic
relations among the charged fields from $\B^L$ and $\B^R$. 
Mathematically, this condition might seem unnecessarily
restrictive. In \cite{BDH}, the notion of defects between chiral
nets is much broader in this respect, admitting degrees of freedom
that are not generated by the given local nets. They are defined only
at the boundary, and cannot be transported by conformal
transformations generated by the stress-energy tensor 
contained in the common joint subnet $\A$. Actually, in their
axiomatization, even trivial nets ($\B^L=\B^R=\CC$) admit many
nontrivial defects \cite[Prop.\ 1.22]{BDH}. The notion of
``additivity'' which is part of the definition of a defect in
\cite{BDH} does not include the condition that the algebra of an
interval across the boundary is generated by the algebras of the two
subintervals on either side. It has the advantage that it admits a
notion of ``fusion'' under which the additive defects are 
closed. 

It is conjectured in \cite[Remark 1.28 in Version 2]{BDH} that 
by assigning $\D(I)$ to an interval across the boundary, and $\B^L(I)$
resp.\ $\B^R(I)$ to intervals away from the boundary, a defect as in
\dref{d:bc} defines a defect in their sense. We do not see how this
very natural conjecture should fail. Conversely, if one adds
``covariance under a common positive-energy representation'' and
``finite index'' to the definition \cite{BDH}, one should obtain a defect in the
sense of \dref{d:bc}.
\end{remark}

In order to {\em classify} boundary conditions between a given pair 
$\B^L$, $\B^R$ (rather than just axiomatizing them), we must find all
representations on a common Hilbert space with the required
properties. In particular, the corresponding nonlocal nets $\D$ are
not known from the outset. In \pref{p:universal} we shall present a
``universal construction'' $\D$ implementing a (non-factorial) boundary
condition, along with a reducible vacuum representation, in such a way
that every irreducible sub-representation of $\D$ (each containing a
unique vacuum vector) implements an irreducible (hence factorial)
boundary condition (\lref{l:mult1}), and every irreducible boundary
condition is obtained in this way.   

In the case of the two-dimensional Ising model, the local observables
$\B_{\rm 2D}$ without a boundary (\sref{E4}) are given by the chiral observables
$\A_+\otimes \A_-=\Vir(c=\frac12)\otimes \Vir(c=\frac12)$ and the charged fields $\Psit$ and $\Psis$
(corresponding to the primary fields $\eps$ and $\sig$ in \sref{E1}),
intertwining the DHR representations with the vacuum representation as
to be described in \sref{s:charged}. In the presence of the boundary,
the universal algebra $\D_{\rm 2D}$ is generated by two copies
$\B^L_{\rm 2D}$ and $\B^R_{\rm 2D}$ of $\B_{\rm 2D}$, sharing the same
chiral observables $\A_{\rm 2D}$. There are three inequivalent
subrepresentations (= boundary conditions): the first one is just the 
theory without a boundary in which the charged fields on either side
coalesce. In the second, the charged fields differ by a gauge
transformation, namely $\Psi_s$ (the order parameter $\sig$) changes sign at the
boundary. In the third, we have $\Psit^L=-\Psit^R$, whereas $\Psis^L$ 
and $\Psis^R$ remain independent fields (the order and the disorder
parameter $\sig$ and $\mu$ in \eref{dualCR}) (\sref{E5}). The product 
$\Psis^L\cdot\Psis^R$ turns out to split into two new primary fields: 
a left-moving and a right-moving chiral Fermi field. Of course, the 
graded-local Fermi fields cannot be constructed out of the local
observables on either side (\sref{E6}).  

\section{Extensions and Q-systems}
\label{s:Ext} \setcounter{equation}{0}

We review the characterization of extensions of quantum field theories
in terms of Q-sytems \cite{LR95}, in particular, the construction of 
the extended local algebras $\B(O)$ from $\A(O)$ and the Q-system. We
then show that the centre of a Q-system corresponds to a maximal local
intermediate extension, which can be obtained in a simple algebraic way.

\subsection{Q-systems and subfactors}
\label{s:Q-ext}
Let us start with a review of the necessary facts from subfactor theory. 

The local algebras in (conformal) quantum field theory are known to be
type $I\!I\!I$ factors $M$ (see \cite[Sect.\ V.3]{B00}
for a review of the type question). This implies that
every nonzero projection $e\in M$ is equivalent to the identity in the sense
that there is an operator $s\in M$ such that $ss^*=e$, $s^*s=\mathbf{1}_M$. 

This feature allows the notions of ``sub-homomorphisms'' and 
``conjugate homomorphisms''. Let $N$ and $M$ by two type $I\!I\!I$
factors, and $\varphi:N\to M$ a homomorphism. (Homomorphisms are always
understood to be unital, i.e., $\varphi(\mathbf{1}_N)=\mathbf{1}_M$.) Let $e\in
\varphi(N)'\cap M$ be a projection in the relative commutant. Let $e=ss^*$
with an isometry $s\in M$. Then
$\varphi_s(\cdot) := s^*\varphi(\cdot)s$
is another homomorphism $\varphi_s:N\to M$. We write $\varphi_s\prec
\varphi$, because if $M$ is represented on a Hilbert space $\H$ and
$N$ is represented on $\H$ via $\varphi$, then $s$ is a unitary
operator from $\H$ carrying the representation $\varphi_s$ to the
subspace $e\H$ carrying the representation $\varphi\vert_{e\H}$. 

The operator $s$ satisfies the intertwining relation
$s\varphi_s(n)=\varphi(n)s$.  
More generally, we call $M\ni t:\varphi_1\to\varphi_2$
an intertwiner between two homomorphisms $\varphi_i:N\to M$, if
$$t\varphi_1(n)=\varphi_2(n)t.$$
The set of intertwiners $t:\varphi_1\to\varphi_2$ is a linear space called
$\Hom(\varphi_1,\varphi_2)\subset M$. If $\varphi_1$ is irreducible
($\Hom(\varphi_1,\varphi_1)\equiv \varphi_2(N)'\cap M=\CC\cdot\mathbf{1}_M$), 
then $t^*t\in\Hom(\varphi_1,\varphi_1)$ is a multiple of $\mathbf{1}_M$, thus
turning $\Hom(\varphi_1,\varphi_2)$ into a Hilbert space (every element
is a multiple of an isometry).  

The direct sum of two homomorphisms $\varphi_i:N\to M$ is defined up to 
unitary equivalence as
$$\varphi(\cdot):=s_1\varphi_1(\cdot)s_1^*+s_2\varphi_2(\cdot)s_2^*,$$ 
with any pair of isometries such that $s_1s_1^*+s_2s_2^*=\mathbf{1}_M$.
Conversely, the decomposition of $\mathbf{1}_M$ into minimal projections in 
$\varphi(N)'\cap M$ gives rise to a decomposition of $\varphi$ as a
direct sum of irreducibles. 

A pair of homomorphisms $\varphi:N\to M$ and $\ol\varphi:M\to N$ is
called conjugate, if $\id_N\prec \ol\varphi\varphi$ and 
$\id_M\prec \varphi\ol\varphi$, and if there are intertwiners $N\ni w:
\id_N\to \ol\varphi\varphi$, and $M\ni v:\id_M\to \varphi\ol\varphi$
satisfying the conjugacy relations 
$$v^*\varphi(w)=\mathbf{1}_M,\quad w^*\ol\varphi(v)=\mathbf{1}_N.$$
One may normalize the pair such that $w^*w=d\cdot \mathbf{1}_N$ and
$v^*v=d\cdot \mathbf{1}_M$. Then the infimum of $d$ among all solutions to the
normalized conjugacy relations is called the {\em dimension}
$\dim(\varphi)=\dim(\ol\varphi)\geq 1$. If a solution of the conjugacy
relations exists at all, then there exist solutions taking the
infimum value. Such solutions are called {\em standard pairs}. 
(Any two standard solutions differ by a pair of unitaries  
$\in \Hom(\varphi,\varphi)$ and $\in \Hom(\ol\varphi,\ol\varphi)$.)
If an endomorphism does not have a conjugate, one puts 
$\dim(\varphi):=\infty$.

The notion of conjugacy is stable under unitary equivalence. More
remarkably \cite{LRo}, the dimension is multiplicative under the 
composition
$$\dim(\varphi_2\circ\varphi_1) = \dim(\varphi_2)\cdot\dim(\varphi_1)$$
and additive under direct sums 
$$\dim(\varphi_1\oplus\varphi_2) = \dim(\varphi_1) + \dim(\varphi_2).$$

If $N\subset M$ is a (type $I\!I\!I$) subfactor, denote
by $\iota:N\to M$ the embedding homomorphism. Then the square of 
$\dim(\iota)$ is equal to the index $[M:N]$ (the type
$I\!I\!I$ version \cite{Ko,KL} of the type $I\!I$ Jones index).
Thus, for finite-index subfactors, there exists a conjugate homomorphism
$\ol\iota:M\to N$ with standard pair $(w,v)$. It also follows that
$[N:\ol\iota(M)]$ has the same index.

Standard pairs give rise to bijective Frobenius maps
$\Hom(\varphi_1,\varphi_2\varphi_3)\to
\Hom(\ol\varphi_2\varphi_1,\varphi_3)$
and $\Hom(\varphi_1,\varphi_2\varphi_3)\to
\Hom(\varphi_1\ol\varphi_3,\varphi_2)$, as well as to positive maps 
$\Hom(\varphi\varphi_1,\varphi\varphi_2)\to \Hom(\varphi_1,\varphi_2)$
$\Hom(\varphi_1\varphi,\varphi_2\varphi)\to \Hom(\varphi_1,\varphi_2)$
with tracial properties \cite{LRo,BKLR}, which we shall freely use in the
rest of the paper. 

These structures turn the homomorphisms between type $I\!I\!I$ factors
into a C* tensor 2-category (with objects the factors, 
1-morphisms the homomorphisms, and 2-morphisms the intertwiners). 
It is again simple and strict, the homomorphisms admit direct sums, and
non-zero projections correspond to sub-homomorphisms.
The endomorphisms of a given factor $N$ form just a C* tensor category
$\End(N)$ (with objects the endomorphisms, and morphisms the
intertwiners). We denote the full subcategory of endomorphisms of finite
dimension as $\End_0(N)$.  

For many purposes below, 
$$\C\subset\End_0(N)$$ 
will stand for any full subcategory of $\End_0(N)$ that without loss
of generality can be assumed to have direct sums and subobjects. 
For more specific results, $\C$ may be assumed to be braided (which is
automatically the case for the representation category of local
quantum field theory, \cite{DHR,FRS}, cf.\ \sref{s:QFT}), and even modular 
(which is automatically the case for completely rational conformal 
chiral quantum field theory \cite{KLM}). 

\medskip

Conjugates and the dimension of $\varphi:N\to M$ can also be defined when  
the condition that $M$ is a factor, is relaxed and one admits 
$M=\bigoplus_i M_i$ to be a finite direct sum of factors. In this case, 
$\varphi(n) = \bigoplus_i\varphi_i(n)$, and $\ol\varphi(\bigoplus_im_i)=
\sum_is_i\ol\varphi_i(m_i)s_i^*$, where $s_i\in N$ are a complete system of 
orthonormal isometries as before. It turns out \cite[Sect.\ 2.3]{BKLR} that the 
dimension $\dim(\varphi)$ (defined as an infimum over solutions of 
the conjugate relations as before) is no longer additive, but
$$\dim(\varphi) = \big(\sum_i\dim(\varphi_i)^2\big)^{1/2}.$$

For the embedding homomorphisms $\iota:N\to M$, one calls 
$\gamma:=\iota\circ\ol\iota:M\to M$ the canonical
endomorphism of $M$, and $\theta:=\ol\iota\circ\iota:N\to N$ the dual
canonical endomorphism of $N$. By definition, $w\in\Hom(\id_N,\theta)$ and
$x:=\ol\iota(v)\in\Hom(\theta,\theta^2)$, and these intertwiners
satisfy 
\be \label{Q-system}
w^*x = \theta(w^*)x=\mathbf{1}_N,\quad xx=\theta(x)x,\quad 
xx^*=\theta(x^*)x,\quad w^*w=x^*x=d\cdot \mathbf{1}_N,
\ee
where $d=\dim(\iota)=\sqrt{\dim(\theta)}$. The first relation is
referred to as {\bf unit property}, the second as {\bf associativity}, the
third as {\bf Frobenius property}, and the last as {\bf standardness}. 
$\iota$ is irreducible ($N'\cap M=\CC\cdot \mathbf{1}$), iff $\id_N$ is
contained in $\theta$ with multiplicity 1. 

This system of equations only refers to the algebra $N$ and its endomorphism
$\theta$. The triple 
\be\label{Qtriple}
A=(\theta,w,x) \qquad
(\theta\in\End(N),w:\id_N\to\theta,x:\theta\to\theta^2)
\ee
is called the Q-system associated with the subfactor $N\subset M$, and
the number $d=\dim(\iota)$ is called the dimension $d_A$ of the
Q-system. 
 
The basic result for our purposes is that the Q-system, i.e., the
triple \eref{Qtriple} satisfying \eref{Q-system} with $d\equiv
d_A=\sqrt{\dim(\theta)}$, allows to reconstruct the subfactor
$N\subset M$ up to isomorphism:  

\begin{theorem} \label{SF-Q} \cite{L94} 
  Let $N$ be a type $I\!I\!I$ 
  factor, and $A=(\theta,w,x)$ a Q-system in $\End_0(N)$ of dimension
  $d_A$, such that $\id_N\prec\theta$ with multiplicity 1. Then there 
  is a (unique up to isomorphism) type $I\!I\!I$ factor $M$ and an 
  irreducible embedding $\iota:N\to M$ such that $A$ is the Q-system 
  associated with this subfactor. The dimension $\dim(\iota)$ equals 
  the dimension $d_A$. 
\end{theorem}

The construction of $M$ from the given data is rather simple. Namely,
$M$ is the algebra generated by $N$ (regarded as a subalgebra written
as $\iota(N)\subset M$) and a single generator
$v\in M$, with the defining relations 
\be \label{SF-rec}
v^* = \iota(w^*x^*)v, \qquad vv = \iota(x)v, \qquad \iota(w^*)v = \mathbf{1}_M, 
\qquad v\iota(n)=\iota(\theta(n))v \quad (n\in N).
\ee
Obviously, every element of $M$ is of the form $\iota(n)v$ with $n\in
N$, and $\iota(n)v=0$ implies $n=0$. 
 
The conjugate homomorphism $\ol\iota:M\to N$ is given by 
\bea\label{iotabar}
\ol\iota:\iota(n)v\mapsto \theta(n)x,
\eea
and $(w,v)$ is a standard solution of the conjugacy relations. The
relative commutant $\iota(N)'\cap M$ is given by elements of the 
form $\iota(q)v$ with $q\in\Hom(\theta,\id)=\CC\cdot w^*$ by assumption,
hence $\iota(N)'\cap M=\CC\cdot \mathbf{1}_M$. Thus, the inclusion is
irreducible, and in particular $M$ is a factor, $M'\cap M=\CC\cdot
\mathbf{1}_M$. Because $M$ is finitely generated from $\iota(N)$, it is closed
in the weak topology inherited from $N$. One can also show the type
$I\!I\!I$ property. 

Notice that a braiding is not required in this result.

\subsection{Decompositions of Q-systems}
\label{s:decoQ}
One can easily generalize \tref{SF-Q} by relaxing the condition 
that $\Hom(\id_N,\theta)$ is one-dimensional, see \cite[Sect.\ 2.3]{BKLR}. 
In this case, the irreducibility $N'\cap M=\CC$ will fail, and $M$ may
even fail to be a factor. 

\begin{theorem} \label{t:vN-Q} Let $N$ be a type $I\!I\!I$ 
  factor, and $A=(\theta,w,x)$ a Q-system in $\End_0(N)$ of dimension
  $d_A$. Then there is a (unique up to isomorphism) type $I\!I\!I$ 
  von Neumann algebra $M$ and an embedding $\iota:N\to M$ with
  conjugate $\ol\iota:M\to N$ such that $A$ is the Q-system associated
  with this conjugate pair. The dimension $\dim(\iota)=\sqrt{\dim(\theta)}$ 
  equals the dimension $d_A$. 
\end{theorem}

If $\dim\Hom(\id_N,\theta)>1$, the inclusion $\iota(N)\subset M$ 
reconstructed from the Q-system as before fails to be irreducible, and the 
von Neumann algebra $M$ may fail to be a factor. Namely $\iota(N)'\cap M$ 
is given by elements of the form $\iota(q)v$ with $q\in
\Hom(\theta,\id_N)$. These are central in $M$ if in addition 
$qx=\theta(q)x$ holds. Correspondingly, we may call a Q-system
{\bf factorial} if $qx=\theta(q)x$ for $q\in\Hom(\theta,\id)$ implies
that $q$ is a multiple of $w^*$. This is the same as ``{\bf simple}'' in the
literature on Frobenius algebras, cf.\ \cite[Sect.\ 3.7]{BKLR}.

In \cite[Sect.\ 4]{BKLR}, we develop different decompositions 
of Q-systems pertinent to these cases. 

More precisely, certain projections in $\Hom(\theta,\theta)$ give rise 
to {\bf reduced Q-systems}. We characterize the precise properties of 
these projections which correspond (i) to the central decomposition of 
$M$ when $M$ is not a factor, (ii) to the irreducible decomposition when 
$M$ is a factor but $N'\cap M\neq \CC$, and (iii) to intermediate 
inclusions $N\subset L\subset M$.

If we have two subfactors $\iota_1(N)\subset M_1$ and $\iota_2(N)\subset M_2$
such that there is an isomorphism $\alpha:M_2\to M_1$, we can consider a new inclusion
$\iota(N)\subset M$ equivalent to each of the inclusions
$\jmath_i(N)\subset M_i$ ($i=1,2$), where $\jmath_1\simeq\iota_1 \oplus
\alpha\circ\iota_2$ and $\jmath_2\simeq\alpha^{-1}\circ\iota_1\oplus\iota_2$. 
Clearly, the Q-systems for $\jmath_i$ are reducible but factorial. 
If $\theta_i=\ol\iota_i\iota_i\in \C$, then the Q-systems for
$\jmath_i$ may not belong to the same sub-category of $\End(N)$, unless 
$\alpha$ can be chosen such that $\ol\iota_1\circ\alpha\circ\iota_2
\in \C$. In this case, we call $\iota_1$ and $\iota_2$ \textbf{Morita
  equivalent} as Q-systems in $\C$. It can easily be checked 
that this is equivalent with the categorical definition of Morita
equivalence of Frobenius algebras using bimodules \cite{KR}. We conclude:

\begin{theorem}
\label{t:moritadirectsum}
Morita equivalent factorial Q-systems in $\C$ can be added to obtain a
factorial Q-system in $\C$, and every factorial Q-system in $\C$ is an
irreducible decomposition of Morita equivalent Q-systems. 
\end{theorem}

\subsection{Q-systems in QFT and extensions}
\label{s:QFT}

We now ``transfer'' the basic result \tref{SF-Q} to quantum field theory
\cite{LR95}. Let $\A$ be a Haag dual net of local von Neumann algebras
$\A(O)$, and $\DHR(\A)$ the category of its DHR endomorphisms. The net
may be two-dimensional or chiral; in the latter case ``$O$'' is
understood to stand for an interval $\subset\RR$.

The transfer from subfactor theory to quantum field theory is possible, 
essentially because if $\rho$ is a DHR endomorphism localized
in $O$, then $\rho$ maps $\A(O)$ into itself, and $\rho(\A(O))\subset
\A(O)$ is a type $I\!I\!I$  subfactor. Likewise, for an extension
$\A\subset\B$, every local inclusion $\A(O)\subset\B(O)$ is a
subfactor. 

DHR endomorphisms localized in $O$, when restricted to $\A(O)$, are in
fact endomorphisms of $\A(O)$, and they have the same intertwiners 
as endomorphisms of the net and as elements of $\End(\A(O))$ 
\cite{GL96}. Therefore, they are the objects of a C* tensor category
$\DHR(\A)\vert_O$, which is a full subcategory of $\DHR(\A)$ and of
$\End(N)$, $N=\A(O)$.  

In other words, $(\theta,w,x)$ is a Q-system in $\DHR(\A)\vert_O$ as
subcategory of $\End(N)$ if and only if $\theta$ is the restriction of
a DHR endomorphism (also denoted by $\theta$) localized in $O$, and 
$w,x\in\A$ satisfy the relations \eref{Q-system}. It is therefore 
safe to drop the distinction altogether. 

Since $\dim(\rho)$ was defined in terms of intertwiners, one may
assign the same dimension to $\rho$ as a DHR endomorphism, and the
same properties (additivity and multiplicativity) remain valid. This  
definition coincides \cite{L89} with the ``statistical dimension'' 
originally defined in terms of the statistics operators \cite{DHR,FRS}.

In Qft models where DHR endomorphisms with infinite dimension occur, we
have to restrict ourselves to the full subcategory $\DHR_0(\A)$ of DHR
endomorphisms with finite dimension. 

Let now $A=(\theta,w,x)$ be a Q-system in $\DHR_0(\A)$, i.e., $\theta$ is a 
DHR endomorphism of $\A$, and the intertwiners $w$ and $x$ are
elements of $\A(O)$ if $\theta$ is localized in $\A(O)$.  

Define an algebra $\B$ by adding one generator $v$ with the relations 
\be \label{QFT-rec}
v^* = \iota(w^*x^*)v, \qquad vv = \iota(x)v, \qquad \iota(w^*)v = \mathbf{1}_\B, 
\qquad v\iota(a)=\iota(\theta(a))v \quad (a\in \A).
\ee
Again, every element is of the form $\iota(a)v$ with $a\in\A$, and
$\iota(a)v=0$ implies $a=0$. 

The local subalgebras of $\B$ are defined as 
\be\label{BO=AOv}\B(O) := \iota(\A(O))\cdot \wh v
\ee
where $\wh v=\iota(u)v$ with $u\in\A$ unitary such that 
$\wh\theta=\Ad_u\theta$ is localized in $O$. Because 
$\wh w=uw\in\Hom(\id,\wh\theta)$ is in $\A(O)$ and
$\iota(\wh w^*)\wh v =\iota(w^*)v=\mathbf{1}$, it follows that 
$\A(O)\subset \B(O)$. $\B(O)$ are *-algebras because $(\wh v)^2 =
\iota(u\theta(u)xu^*)\wh v$, 
where $u\theta(u)xu^*\in\Hom(\wh\theta,\wh\theta^2)\subset \A(O)$, and 
$\wh v^*=\iota(w^*x^*\theta(u^*)u^*)\wh v$, where
$u\theta(u)xw\in\Hom(\id,\wh\theta^2)\subset \A(O)$.  

One has 
$$\iota(u)v\iota(a) = \iota\big(u\theta(a)\big)v =
\iota\big(\wh\theta(a)\big)\wh v=\iota(a)\wh v$$
whenever $a\in\A(O')$, hence the extended net $\B$ is relatively local
w.r.t.\ $\A$. Moreover, $\B$ is itself local if $\iota(u_i)v$
localized in spacelike separated $O_i$ commute, which turns out to be
equivalent to the condition 
$$\eps_{\theta,\theta}\, x = x.$$

The conjugate homomorphism $\ol\iota$ is again defined by \eref{iotabar}.
There is a conditional expectation $\mu:\B\to\A$ (i.e., a faithful
positive unit-preserving linear map satisfying 
$\mu(\iota(a_1)\cdot b\cdot\iota(a_2))=a_1\cdot\mu(b)\cdot a_2$,
generalizing the properties of an average over an automorphic group
action), defined by  
\be\label{condex}
\mu(b)=d_A\inv\cdot w^*\ol\iota(b)w.
\ee
The conditional expectation respects the local structure, namely
$\mu(\B(O))=\A(O)$. 

With the help of the conditional expectation \eref{condex}, the vacuum
state $\omega_0$ of $\A$ extends to a vacuum state
$\omega^0:=\omega_0\circ\mu$ of $\B$. The GNS representation of the
restriction $\omega^0\vert_\A$ as a state on $\A$ is unitarily
equivalent to the representation $\pi_\theta=\pi_0\circ\theta$ given
by the DHR endomorphism $\theta$. 

The upshot of this discussion is 

\begin{theorem} \label{t:QFT-Q} \cite{LR95} Let $\A$ be a (conformal)
  QFT, and $A=(\theta,w,x)$ a Q-system in $\DHR_0(\A)$ of dimension
  $d_A$. Then there is
  a (unique up to isomorphism) QFT $\B$ with local algebras
  \eref{BO=AOv} and an irreducible embedding
  $\iota:\A\to \B$ such that $A$ is the Q-system associated with this
  extension. The dimension $\dim(\iota)$ equals the dimension $d_A$. 
  $\B$ is relatively local w.r.t.\ $\A$, and $\B$ is local iff
\be\label{local} 
\eps_{\theta,\theta}\, x = x
\ee
  holds. The conditional expectation
  $\mu:\B\to\A$ extends the vacuum state on $\A$ to a vacuum state on
  $\B$, such that the GNS representation of its restriction to $\A$ is
  equivalent to $\theta$. 
\end{theorem}

Here, $\DHR_0(\A)$ is the full braided subcategory of DHR endomorphisms
of finite dimension. Notice that the objects of $\DHR_0(\A)$ are just
the finitely reducible objects of $\DHR(\A)$ if $\A$ is
completely rational. For a more ``physical'' interpretation of the
extension, see the \sref{s:charged} below.

\medskip

\tref{t:QFT-Q} was proven in \cite{LR95} under the additional assumption that 
$\dim\Hom(\id_\A,\theta)=1$, in which case the extension is irreducible, 
and the local algebras are factors. As in \tref{t:vN-Q}, this condition 
can be relaxed without difficulty, thus admitting also reducible
extensions and extensions whose local algebras are not factors. This
generalization is necessary because in the analysis of 
boundary conditions, it will turn out that one has 
to consider Q-systems that give rise to extensions with a nontrivial 
centre, $\B'\cap\B\not=\CC\cdot \mathbf{1}_\B$.  

The central decomposition of $\B$ will give rise to inequivalent
representations (and inequivalent boundary conditions). It is
therefore necessary to characterize the centre of the extension, and
its central projections.  

\begin{lemma} \label{l:center} {\rm (i)} (With assumptions as in
  \tref{t:vN-Q}, admitting $\id\prec\theta$ with multiplicity $>1$)
  The relative commutant is given by $\iota(N)'\cap
  M=\iota(\Hom(\theta,\id_N))v$. The centre of $M$ is given by
  $\iota(q)v$, $q\in\Hom(\theta,\id_N)$ satisfying 
\be \label{centre-char} qx=\theta(q)x.\ee  
  $\iota(q)v$ is selfadjoint iff $q^*=\theta(q)x w$, and
  idempotent iff $q\theta(q)x\equiv qqx=q$. 
\\[1mm]
  {\rm (ii)} (With assumptions as in \tref{t:QFT-Q}, admitting
  $\id\prec\theta$ with multiplicity $>1$) The
  relative commutant is given by $\iota(\A)'\cap
  \B=\iota(\Hom(\theta,\id_\B))v$. The centre of $\B$ is given by
  $\iota(q)v$, $q\in\Hom(\theta,\id_\B)$ satisfying
  \eref{centre-char}. $\iota(q)v$ is selfadjoint iff
  $q^*=\theta(q)xw$, and idempotent iff $q\theta(q)x\equiv qqx=q$. 
\end{lemma}

The proof proceeds by direct computation, using the relations
\eref{SF-rec}, \eref{QFT-rec}.

\subsection{Charged fields}
\label{s:charged}

The reconstruction presented above by adjoining an element $v$ to the
algebra $\A$ of local observables and imposing its relations, can be 
given a physical interpretation. Namely, the DHR endomorphism $\theta$ 
gives the restriction of the vacuum representation of the larger net 
$\B$ as a representation of the original net $\A$. If $\rho\prec\theta$ 
is an irreducible subrepresentation localized in $O$, then 
$H_\rho:=\iota(\Hom(\rho,\theta))^*v$ is a Hilbert 
space of isometries in $\B(O)$, and its elements 
$\psi_\rho=\iota(w_\rho)^*v\in \B(O)$ with
$w_\rho\in\Hom(\rho,\theta)$ satisfy the commutation relation 
\be\label{charged}
\psi_\rho a = \rho(a) \psi_\rho
\ee
with the observables in $\A$, intertwining the vacuum representation
with the representation $\rho$. Because $\rho$ is localized in $O$, it
follows that $\psi_\rho$ commutes with $\A(O')$, i.e., the extension
is relatively local w.r.t.\ $\A$. 

Because $\mu(\psi_\rho\psi_\rho^*)\in\Hom(\rho,\rho)=\CC\cdot\mathbf{1}$,
one may normalize $\psi_\rho$ such that $\mu(\psi_\rho\psi_\rho^*)=\mathbf{1}$. 
Then $\omega_\rho(a)=\omega^0(\psi_\rho a \psi_\rho^*)\equiv 
\omega_0\circ\mu(\psi_\rho a \psi_\rho^*)$ is a state on $\A$. 
Its GNS representation is equivalent to the representation 
$\pi_\rho=\pi_0\circ\rho$, justifying the interpretation of
$\psi_\rho$ as ``charged field operators''. 

By choosing a complete irreducible decomposition of $\theta$ by
isometries $w_\rho$ (possibly with multiplicities) such that 
$\sum_\rho w_\rho w_\rho^*=\mathbf{1}$, every element of $\B$ has a unique
representation
$$b = \iota(a)v = \sum_\rho \iota(a_\rho)\cdot\psi_\rho$$
with $a=d_A\cdot \mu(bv^*) = w^*\ol\iota(b)$ and $a_\rho
  =aw_\rho\in\A$, 
so that the operators $\psi_\rho$ form a basis of $\B$ as an
$\A$-module. Likewise, if $\psi_\rho\in\B(O)$, they form a basis of
$\B(O)$ as an $\A(O)$-module. These bounded operators may therefore
be regarded as the analogue of the primary fields in AQFT, such that a
general (``descendant'') field is obtained by left multiplication
(``OPE'') of $\psi_\rho$ with some element of $\A$. 

The formulae given for the product $v^2$ and the adjoint $v^*$
determine the multiplication and adjoint of the charged field
operators $\psi_\rho$. In other words, the Q-system is a 
generating functional for the coefficients of the operator product
expansion of the charged field operators. 

In four dimensions, the Doplicher-Roberts reconstruction \cite{DR}
theorem states (among other things) that for every DHR endomorphism
$\rho$ there is an orthonormal basis of charged field operators
satisfying the Cuntz relation $\sum_i\psi_{\rho,i}\psi_{\rho,i}^*=\mathbf{1}$,
which implement $\rho$ by the formula $\rho(a)=\sum_i\psi_{\rho,i} a
\psi_{\rho,i}^*$. 

In low dimensions, the non-triviality of the braiding poses an
obstruction, so that the range projections of $H_\rho$ in general do
not exhaust the unit operator; but one has still the implementation
via the conditional expectation: $\rho(a)=\mu(\psi_\rho a\psi_\rho^*)$, 
assuming that
$\mu(\psi_\rho \psi_\rho^*)\in\Hom(\rho,\rho)=\CC\cdot\mathbf{1}$ 
is normalized to 1. More details can be found, e.g., in \cite{LR04}.

Let us for later use compute the proper normalization of the charged
fields. We have 
$$\mu(\psi_\rho\psi_\rho^*) = d_A\inv\cdot
w^*\ol\iota(\psi_\rho\psi^*_\rho)w = d_A\inv\cdot
w^*\theta(w_\rho^*)xx^*\theta(w_\rho)w = d_A\inv\cdot
w_\rho^*w^*xx^* ww_\rho = d_A\inv\cdot w_\rho^*w_\rho.$$
Thus, $\mu(\psi_\rho\psi_\rho^*)=\mathbf{1}$ if $w_\rho^*w_\rho=d_A$. Later, we
shall prefer a normalization such that $\psi_\rho$ are
isometries. This is a reasonable option if $\iota$ is irreducible
because $\psi_\rho^*\psi_\rho\in\A(O)'\cap \B(O)$ by \eref{charged},
hence is automatically a multiple of $\mathbf{1}$. 

In this case, $\psi_\rho^*\psi_\rho$ equals its image under $\mu$. 
We compute $\mu(\psi_\rho^*\psi_\rho)=d_A\inv\cdot w^*x^*\theta(w_\rho
w_\rho^*)xw =d_A\inv\cdot \ol w_\rho^* w_\rho^*
w_\rho \ol w_\rho$ by the properties of standard pairs
\cite{LRo}. Because $\ol w_\rho^* \ol w_\rho=\dim(\rho)$, we conclude
that, in the irreducible case, $\psi_\rho=\iota(w_\rho)^*v$ are
isometries iff
\be\label{wnorm} w_\rho^*w_\rho=\frac{d_A}{\dim(\rho)}.
\ee
Similarly, if $\dim\Hom(\rho,\theta)>1$, then
$\psi_\rho^n:=\iota(w_\rho^n)^*v$, $n=1,\dots\dim\Hom(\rho,\theta)$,
are mutually orthonormal isometries iff $w_\rho^n\in\Hom(\rho,\theta)$
are normalized as $w_\rho^{n*}w_\rho^{m}=\delta_{nm}\cdot d_A/\dim(\rho)$.

Charged fields $\psi_\rho$ are localized in $\B(O)$ if $\rho$ is
localized in $O$. They can be transported to any other region
$O_1$ with the help of unitary intertwiners $u\in\Hom(\rho,\rho_1)$
where $\rho_1$ is localized in $O_1$, hence $w_{\rho_1} = w_\rho
u^*\in\Hom(\rho_1,\theta)$: 
\be\label{transport}
\psi_{\rho_1} = \iota(w_{\rho_1})^*v = \iota(u)\psi_\rho.
\ee

\subsection{Example 3: The fermionic extension of the chiral Ising
  model} \label{E3}

The irreducible Q-systems in a given tensor category can be
  computed by making an ansatz for $\theta$ as a sector (exploiting the upper
  bound for the multiplicities
  $\dim\Hom(\rho,\theta)\leq\dim(\rho)$) \cite{LR95}), and solving 
the defining relations for $w\in\Hom(\id,\theta)$ and
$x\in\Hom(\theta,\theta^2)$ in finite-dimensional intertwiner spaces.
For the chiral Ising model (= $\Vir(c=\frac12)$), we use the formulae
from Example 2, \sref{E2}, and just present the results.

Apart from the trivial Q-system $(\id,1,1)$ with $\B=\A$, there is one
other irreducible Q-system with $\theta\sim\id\oplus\tau$ of dimension
$d_A=\sqrt2$. Up to unitary equivalence, we may choose
$\theta=\sigma^2$ and get $w=2^{1/4} \cdot r$,
$x=2^{1/4}\cdot\sigma(r)=2^{-1/4}\cdot (r+t)$. The nontrivial charged 
field is $\psi=2^{1/4}\cdot \iota(t^*)v$. By \eref{SF-rec}, it satisfies
$\psi \iota(a):=\iota(\tau(a))\psi$, $\psi^*=\psi$, $\psi^2=\mathbf{1}$.

If $\tau$ is localized in $I$, $\psi\in\B(I)$. With $\alpha_x$ the
translation automorphism, $\tau_x=\alpha_x\tau\alpha_{-x}$ is
localized in $I+x$. With a unitary charge transporter
$u_{x}\in\Hom(\tau,\tau_{x})$, one can shift  
the localization of $\psi$, putting $\psi_x:=u_{x}\psi\in\B(I+x)$.
Then $\psi_x\psi_y=u_x\tau(u_y)$, and in particular, the local 
anti-commutativity $\psi_x\psi_y=-\psi_y\psi_x$ follows whenever the 
localization regions $I+x$ and $I+y$ are disjoint, because 
$\tau(u_x)^*u_y^*u_x\tau(u_y)$ equals the statistics operator 
$\eps_{\tau,\tau}^\pm=-1$. Indeed, the unitary operator $\psi$ is the
real chiral Fermi field smeared with a test function supported in $I$. 

The extension $\B$ is graded local, with grading automorphism
$\alpha:\psi\mapsto-\psi$, such that $\A=\B^\alpha$, i.e., $\alpha$ is
the global gauge transformation with fixed points $\A$.

\subsection{The canonical extension} \label{s:canonical}
In two dimensions, the same results can be applied to 
$\A_{\rm 2D}=\A_+\otimes\A_-$ with
$\DHR_0(\A_+\otimes\A_-)=\DHR_0(\A_+)\boxtimes\DHR_0(\A_-)\opp$. In order to
distinguish it from the chiral case, we adopt capital letters 
for the two-dimensional case. 

A Q-system is a triple $(\Theta,W,X)$, where $\Theta$ has the general form 
\be\label{Z}
\Theta \simeq \bigoplus_{\rho,\sig}Z_{\sig,\tau} \cdot\sig\otimes\tau.
\ee
The multiplicity matrix $Z$, coupling the left- and the right-moving
representations, is also called the {\bf coupling matrix}. 

Thus, from a Q-system $(\Theta,W,X)$, one can construct an extension 
$\A_{\rm 2D}\subset\B_{\rm 2D}=\iota(\A_{\rm 2D})V$. The charged fields are of the form
$\Psi^i_{\sig\otimes\tau}=\iota(W^i_{\sig\otimes\tau})^*V$
satisfying 
$$\Psi^i_{\sig\otimes\tau}a =
(\sig\otimes\tau)(a)\Psi^i_{\sig\otimes\tau}
$$
where $\sigma\otimes\tau\prec\Theta$ are the irreducible
subrepresentations of $\Theta$, and
$i=1,\dots Z_{\sig,\tau}$. 

In the case $\A_+=\A_-$, there is a distinguished extension of
$\A_{\rm 2D}=\A\otimes\A$ whose coupling matrix is the conjugation matrix.

\begin{proposition}\label{p:canonical}
 \cite{LR95} If $\DHR_0(\A)$ has finitely 
  many inequivalent irreducible objects $\rho$, then there is a 
  {\bf canonical Q-system} $R$ in $\DHR_0(\A\otimes\A)$ with
  $$\Theta\simeq\bigoplus\rho\otimes\ol\rho,$$ 
  where the sum extends over the irreducible sectors of 
  $\A$. More precisely, choosing isometries
  $T_\rho\in\Hom(\rho\otimes\ol\rho,\Theta)$, one has
  $W=d_R^{1/2}\cdot T_\id$ and 
$$X=d_R^{-1/2}\sum_{\rho,\sig,\tau}
\Big(\frac{d_\rho d_\sig}{d_\tau}\Big)^{1/2}\cdot\Theta(T_\sig)T_\rho
\circ \Big(\sum_a t_a\otimes t_{\ol a}\Big) \circ T_\tau^*,$$
where the second sum extends over an orthonormal basis of isometries
$t_a\in\Hom(\tau,\rho\sig)$ and 
$t_{\ol a}\in\Hom(\ol\tau,\ol\rho\ol\sig)=j(t_a)$ with a suitable
antilinear conjugation such that $\ol\rho=j\circ\rho\circ j$ for all
representatives. The dimension is $d_R=\sqrt{\dim(\Theta\can)}=
\sqrt{\sum_\rho \dim(\rho)^2}$. 
\end{proposition}

The charged operators $\Psi_{\rho\otimes\ol\rho}$ of the
  canonical extension are isometries if
$W_{\rho\otimes\ol\rho}\in\Hom(\rho\otimes\ol\rho,\Theta)$ are normalized as
$W_{\rho\otimes\ol\rho}^*W_{\rho\otimes\ol\rho}=\sqrt{\dim(\Theta)}/\dim(\rho)^2$
in accordance with \eref{wnorm}.

The canonical Q-system can be defined in more general settings, e.g.,
as a Frobenius algebra in $\C \boxtimes \overline{\C}$ where
$\C$ is a ribbon category and $\overline{\C}$ its dual (with inverted
morphisms) \cite[Lemma 6.19]{KR}, or in tensor
categories without a braiding, replacing the
conjugate by $\rho^j=j\circ\rho\circ j$ \cite[Prop.\ 4.10]{LR95}. It is also
referred to as {\bf regular} because it
shares certain properties with the regular representation of a group. 
In other contexts, the corresponding extensions are also knwon as 
``Cardy extension'' or ``Longo-Rehren subfactor''. 

We have
\begin{proposition} \cite{LR95} The canonical Q-system satisfies 
$$\eps_{\Theta,\Theta}\cdot X=X,$$
i.e., it is commutative in $\DHR_0(\A\otimes\A)$.
\end{proposition}

Thus, the
canonical Q-system describes a local two-dimensional extension
$\A\otimes\A\subset \B_{\rm 2D}$ with charged fields $\Psi_{\rho\otimes\ol\rho}$.

\subsection{Example 4: The local two-dimensional Ising model}
\label{E4}

The Ising model (without phase boundary) is a local extension of the theory
$\A_{\rm 2D}=\A\otimes \A$, where $\A$ is the Virasoro theory with $c=\frac12$.
The Q-system is the canonical Q-system (\pref{s:Ext}.5)
given as follows. Let $\tau$ and $\sigma$ be localized in $I$, and
$O=I\times I$. Choose {\em any} three isometries $T_0$, $T_1$, $T_2\in
(\A\otimes\A)(O)$ satisfying $T_i^*T_j=\delta_{ij}\cdot\mathbf{1}$,
$\sum_{i}T_iT_i^*=\mathbf{1}$, and define 
$$\Theta(a)=T_0aT_0^* + T_1(\tau\otimes\tau)(a)T_1^* +
T_2(\sig\otimes\sig)(a)T_2^*.$$ 
The intertwiners of the canonical Q-system are given by $W:=\sqrt 2
\cdot T_0$ and \cite{LR95}  
\bea
X=\Big(\Theta(T_0)T_0T_0^*+ \Theta(T_0)T_1T_1^*+ \Theta(T_0)T_2T_2^*+
\Theta(T_1)T_0T_1^*+ \Theta(T_2)T_0T_2^*+
\Theta(T_1)T_1T_0^*+\quad\notag \\[-1mm] 
+ \Theta(T_2)T_1T_2^*+ \Theta(T_1)T_2(u\otimes
u)T_0^*+ \sqrt2\cdot \Theta(T_2)T_2(r\otimes r)T_0^*+
\sqrt2\cdot \Theta(T_2)T_2(t\otimes t)T_1^*\Big)/\sqrt 2.\notag\eea
Properly normalizing $W_{a\otimes a}=\sqrt2/\dim(a)\cdot T_a$,
one obtains isometric operators $\Psit=\iota(W_{\tau\otimes\tau})^*V$
and $\Psis= \iota(W_{\sig\otimes\sig})^*V$ satisfying the relations  
$$\Psit a = (\tau\otimes\tau)(a)\Psit,\quad\Psis a = (\sig\otimes\sig)(a)\Psis$$
$$\Psit^*=\Psit,\quad\Psis^*=\sqrt2 (r^*\otimes r^*)\Psis,$$
$$\Psit^2=\mathbf{1},\quad\Psit\Psis=\Psis,\quad\Psis\Psit=(u\otimes
u)\Psis,$$
$$(\Psis)^2=((r\otimes r)+(t\otimes t)\Psit)/\sqrt2.$$
(The first line holds by construction, the others follow
by direct computation using \eref{SF-rec}, adapted to the case at
hand, and the formulae from Example 2, \sref{E2}.)

Then $\B_{\rm 2D}(I\times I)$ is generated by $\A_{\rm 2D}(I\times I)$ and $\Psit$
and $\Psis$. Shifting the localization by unitary charge transporters,
as in Example 3, \sref{E3}, one can explicitly verify local commutativity among
all fields at spacelike distance.  

One may add also a chiral Fermi field to this algebra, giving rise to
a nonlocal extension $\A_{\rm 2D}\subset \D_{\rm 2D}$ with $\Theta \simeq 
(\id\oplus\tau)\otimes(\id\oplus\tau)\oplus2\sig\otimes\sig$
containing the canonical local extension $\B_{\rm 2D}$ as an intermediate extension.
The relations of this extension will be presented in Example 5, \sref{E5}.

\subsection{Maximal local subtheories of nonlocal extensions}
\label{s:max}

We introduce two maximal local extensions contained in a nonlocal
extension. 

As before, $O$ may be a double cone in two dimensions, or an
interval $\subset\RR$, and $O'$ the causal complement, resp.\ the
complement in the chiral case. In the two-dimensional case, we denote
the connected components of $O'$ as $W_L$ (the ``left wedge'') and
$W_R$ (the ``right wedge''). In the chiral case, the two connected
components are in fact halfrays, but we shall adopt the same
notation $W_L$ (for the negative=past halfray) and $W_R$ (for the
positive=future halfray). For wedge algebras (halfray algebras), Haag
duality holds, namely $\A(W')=\A(W)'$. 

Let $\A\subset\B$ be an extension. We define 
$$\B^+\loc(O):=\B(W_L)'\cap \B(W_R'), \quad\hbox{resp.}\quad 
\B^-\loc(O):=\B(W_R)'\cap \B(W_L').$$
These algebras are the relative commutants of the pair of algebras 
associated with the nested left resp.\ right wedges (past resp.\ future
halfrays) defined by the double cone (interval) $O$. 

\smallskip

\begin{lemma}\label{Ocomm} 
{\rm (i)} $\B\loc^\pm$ are Poincar\'e covariant isotonous nets of von Neumann
algebras. \\
{\rm (ii)} $\B$ is left-local w.r.t.\ $\B\loc^+$, and right-local w.r.t.\
$\B\loc^-$. Equivalently, $\B\loc^+$ is right-local, and $\B\loc^-$ is
left-local w.r.t.\ $\B$. \\ 
{\rm (iii)} The nets $\B\loc^\pm$ are both local, 
and they are intermediate between $\A$ and $\B$:
$$\A(O)\subset\B^\pm\loc(O)\subset \B(O).$$
In particular, one has 
\be \label{Bloc-O}
\B^+\loc(O)=\B(W_L)'\cap \B(O), \quad\hbox{resp.}\quad 
\B^-\loc(O)=\B(W_R)'\cap \B(O),
\ee
where the wedges $W_L$ and $W_R$ are the left and right connected components of
$O'$.  \\
{\rm (iv)} The nets $\B\loc^\pm$ are the maximal intermediate nets
with the properties stated in {\rm (ii)}.
\end{lemma}

\begin{proof} 
(i) Enlarging the double cone $O$, certainly decreases
$W_L$ and/or $W_R$ and increases $W_L'$ and/or $W_R'$. Therefore 
$\B\loc^\pm(O)$ increase with $O$ (``isotony''). The Poincar\'e
covariance of the constructions is manifest. \\
(ii) Clear by definition. \\ 
(iii) To prove locality, it suffices to note that for every pair of 
spacelike separated double cones, one is contained in the left
component of the complement of the other, and the other is contained 
in the right component of the complement of the first. Locality then 
follows because the algebras are defined as relative commutants. 

The inclusion $\A(O)\subset \B\loc^\pm(O)$ is obvious from relative
locality of $\B$ w.r.t.\ $\A$. The inclusion
$\B\loc^\pm(O)\subset\B(O)$ is not immediate because $\B(W_R')$ is
larger than $\B(O)$. 

Consider an element $b$ of $\B^+\loc(O)=\B(W_L)'\cap \B(W_R')$. 
By \eref{BO=AOv}, $\B(W_R')=\A(W_R')\wh v$ where $\wh v$ can be chosen in
$\B(O)\subset \B(W_R')$. Thus $b=a\wh v$ with $a\in\A(W_R')$. Because $b$
commutes with $\B(W_L)$, it commutes with $\A(W_L)$. But $\wh v$
commutes with $\A(W_L)$ by relative locality, hence $a$ must commute
with $\A(W_L)$. Then, $a\in\A(W_L')$ by Haag duality, and hence
$a\in \A(W_R')\cap\A(W_L') = \A(O)$. Thus $b\in \A(O)\wh v = \B(O)$. 
The argument is the same for $\B\loc^-$. 
\\ 
(iv) Is now clear from \eref{Bloc-O}.  
\end{proof}

If $\B$ is itself local, then obviously $\B^+\loc=\B^-\loc=\B$. Also the other
extreme, $\B^\pm\loc=\A$, may occur; e.g., the fermionic extension of the
chiral Ising model (Example 3, \sref{E3}) does not admit any local
intermediate extension.  

Let $\A\subset \B_1\subset\B$ be any intermediate extension
with Q-system $(\theta_1,w_1,x_1)$. Then
$\iota=\iota_2\circ\iota_1$ and $\theta_1 =
\ol\iota_1\iota_1\prec \ol\iota\iota=\theta$. Let
$p\in\Hom(\theta,\theta)$ be the projection corresponding to
$\theta_1\prec\theta$. It is easy to see that 
\be\label{intermediate}
pw=w,\qquad p\theta(p)x = pxp=\theta(p)xp.
\ee
In fact, every projection $p\in\Hom(\theta,\theta)$ satisfying  
\eref{intermediate} comes from an intermediate extension
$$\A \subset \B_p \subset \B$$
where $\B_p = \A\cdot pv$. This is proven in \cite[Sect.\ 4.4]{BKLR} 
for von Neumann algebras, and translates to local nets by 
$\B_p(O):= \A(O)\cdot s^*v$ where $p=ss^*$, $s^*s=\mathbf{1}$, and 
$s^*\theta(\cdot)s$ is localized in $O$.

\begin{lemma}\label{ll} $\B$ is left-local w.r.t.\ $\B_1$ if and only if $p$
satisfies 
\be\label{rightlocal}
\theta(p)x = \eps_{\theta,\theta}\cdot px.
\ee
$\B$ is right-local w.r.t.\ $\B_1$ if and only if $p$
satisfies the same relation with
$\eps_{\theta,\theta}$ replaced by $\eps^*_{\theta,\theta}$. 
\end{lemma}

\begin{proof}
For left locality, we have to consider the commutativity between 
$\B(W_L)$ and $\B_1(O)$ where $O$ is some double cone (interval) and
$W_L$ is the left connected component of $O'$. 
Let $\theta$ be localized in $O$, and choose a unitary
$u\in\Hom(\theta,\wh\theta)$ with $\wh\theta$ localized in $W_L$. Then 
$\B(W_L)=\A(W_L)uv$ as in \eref{BO=AOv}. Let $p=ss^*$ with an isometry
$s\in\A(O)$, hence $\theta_1$ is localized in $O$. Then we also
have $\B_1(O)=\A(O)v_1=\A(O)s^*v=\A(O)s^*pv=\A(O)pv$.

Now, $\A(O)$ commutes with $\B(W_L)$ by relative locality, and
$\A(W_L)$ commutes with $\B_1(O)$ by relative locality. Thus
$\B_1(O)$ commutes with $\B(W_L)$ if and only if $uv$ commutes with $pv$. 
We have 
$$uvpv=u\theta(p)xv,\qquad pvuv = p\theta(u)xv =
\theta(u)pxv.$$
Equality holds iff $\theta(p)x = u^*\theta(u)px$. 
But
$u^*\theta(u)=\eps_{\theta,\theta}$, so the claim follows.
\end{proof}

We now claim that in terms of Q-systems, the above constructions
$\B\loc^\pm$ correspond to the {\bf centres of the Q-system}
\cite[Sect.\ 2.4]{FFRS06}. The centre of a Q-system
$(\theta,w,x)$ (which in general does not satisfy the commutativity
condition $\eps_{\theta,\theta}x=x$) is given by a projection $p$ in
$\Hom(\theta,\theta)$, such that the associated intermediate Q-system
$(\theta_p,w_p,x_p)$ satisfies $\eps_{\theta_p,\theta_p}x_p=x_p$ and
is maximal with this property. There are in fact two such projections,
$p^\pm$, and hence two centres, defined by 
\be\label{centerproj}
\dim(\theta)^{\frac12}\cdot p^\pm:=
r^*\theta(\eps^\pm_{\theta,\theta})x^{(2)} \equiv
\theta(r^*)\eps^\mp_{\theta,\theta}x^{(2)}, 
\ee
where $x^{(2)}=xx=\theta(x)x\in\Hom(\theta,\theta^3)$ and
$r=xw\in\Hom(\id,\theta^2)$. They satisfy the relations 
\eref{intermediate}, and in addition 
\be\label{centerrel}
\theta(p^+)x = \eps_{\theta,\theta}\cdot p^+x, \qquad \theta(p^-)x =
\eps^*_{\theta,\theta}\cdot p^-x,  
\ee
and they are maximal to satisfy \eref{centerrel}, i.e., for every
other projection $p$ satisfying the first resp.\ the second of
\eref{centerrel}, one has $pp^+ = p$ resp.\ $pp^- =
p$. The intermediate Q-system associated with $p^+$ is called 
the {\bf left centre} $C^+[A]$ of $A=(\theta,w,x)$, the Q-system 
associated with $p^-$ is called the {\bf right centre} $C^-[A]$. 

\eref{centerrel} entails $\eps_{\theta_p,\theta_p}x_p=x_p$
\cite{FFRS06}. The proof can also be found in \cite{BKLR}. It follows

\begin{proposition} \label{p:relcomm=cent}
The intermediate extensions associated with the left resp.\ right
centre of the Q-system for $\A\subset\B$ are
$\A\subset\B\loc^+\subset\B$ resp.\ $\A\subset\B\loc^-\subset\B$. 
\end{proposition}

\begin{proof} 
By \lref{Ocomm}, $\B\loc^+$ is intermediate between
$\A\subset\B$ and right-local w.r.t.\ $\B$ $\LRA$ 
$\B$ is left-local w.r.t.\ $\B\loc^+$, and by definition, they are
maximal with these properties. By \lref{ll}, the
associated intermediate projection $p$ is the maximal projection
satisfying \eref{centerrel}. But these properties characterize the 
centre projection, hence $p=p^+$. Similar for $\B\loc^-$. 
\end{proof}

This result gives a simple interpretation of the ``centres of a
Q-system'' in terms of local algebras, namely as the relative
commutant of local algebras associated with nested wedges (in two
dimensions) or lightrays (in chiral theories).

It also provides us a simple formulation of the locality condition
required at a phase boundary: 

\begin{corollary}
\label{c:B<Dloc}
An extension $\D$ of $\A$ implements a defect between $\B^L$ and
$\B^R$, if and only if one has 
$$\B^L\subset \D\loc^-\subset\D\supset\D\loc^+\supset \B^R.$$ 
\end{corollary}

\begin{proof}
By \pref{p:relcomm=cent}, we have $\D\loc^+(O)\subset \D(O)$, hence
$\D\loc^+(O) = \D(O)\cap \D(W_L)'$. Then the defining properties
of a defect, $\B^R(O) \subset \D(O)$ and left-locality of $\D$ w.r.t.\
$\B^R$, are equivalent to $\B^R(O)\subset \D\loc^+(O)$. Similar for $\B^L$. 
\end{proof}

\subsection{The braided product of  two extensions}
\label{s:product}

Suppose we have two extensions $\A\subset\B_1$ and $\A\subset\B_2$
with generating charged fields $\psi_{1,\rho}$ ($\rho\prec\theta_1$) and
$\psi_{2,\sig}$ ($\sig\prec\theta_2$). The two extensions in general
``live'' on different Hilbert spaces, that are both DHR representations of
the underlying algebra $\A$. 

We want to construct an extension $\A\subset\D$ containing 
both $\B_1$ and $\B_2$ as intermediate extensions. Because of the 
commutation relations $\psi_{1,\rho} a=\rho(a)\psi_{1,\rho}$ and 
$\psi_{2,\sig} a=\sig(a)\psi_{2,\sig}$ that must also hold in $\D$,
one may expect algebraic consistency problems. The following
proposition shows that such a construction is canonically possible. 

\begin{proposition} \label{p:productQ}
Let $A_1=(\theta_1,w_1,x_1)$ and $A_2=(\theta_2,w_2,x_2)$ be two Q-systems. 
Then the triples
\be\label{prodQ}
(\theta_1,w_1,x_1)\times^\pm(\theta_2,w_2,x_2) =
(\theta=\theta_1\theta_2,\,w=w_1w_2,\,x^\pm=
\theta_1(\eps^\pm_{\theta_1,\theta_2})x_1\theta_1(x_2))
\ee
define a pair of Q-systems denoted $A_1\times^\pm A_2$.
We call them {\bf (braided) product Q-systems}, and denote the corresponding 
{\bf (braided) product of extensions} by $\A\subset
\B_1\times^\pm\B_2\equiv\D^\pm$. Both product Q-systems $A_1\times^\pm
A_2$ contain $A_1$ and $A_2$ as intermediate Q-systems, hence
$\D^\pm$ contain $\B_1$ and $\B_2$ as intermediate extensions:
$$\begin{array}{ccccc}
&&\B_1&&\\[-1.0mm]
&\rotatebox[origin=c]{30}{$\subset$}&&\rotatebox[origin=c]{-30}{$\subset$}&
\\[-1.0mm] \A&&&&\D^\pm\\[-1.0mm]
&\rotatebox[origin=c]{-30}{$\subset$}&&\rotatebox[origin=c]{30}{$\subset$}&
\\[-1.0mm] &&\B_2&&
\end{array}.$$ 
$\D^\pm$ can be characterized as the quotient of the free product of
the algebras $\B_1$ and $\B_2$ by the relations
$\iota_1(\A)=\iota_2(\A)$ and the commutation relations  
\be\label{comm-v}
v_2v_1 = \eps_{\theta_1,\theta_2}^\pm \cdot v_1v_2.
\ee
\end{proposition}

\begin{proof}
The braided product of Q-systems can be found in 
\cite[Sect.\ 3.2]{FFRS06}, and its defining relations are trivially verified.
The projections $p_1=d_2\inv \theta_1(w_2w_2^*)
\in\Hom(\theta,\theta)$ and $p_2=d_1\inv
w_1w_1^*\in\Hom(\theta,\theta)$ define intermediate Q-systems 
according to \cite[Sect.\ 4.4]{BKLR}. The reduced Q-systems 
coincide with $(\theta_1,w_1,x_1)$ and $(\theta_2,w_2,x_2)$, hence 
both $\B_1$ and $\B_2$ are intermediate extensions with 
$v_1=\theta_1(w_2^*)v$ and $v_2=w_1^*v$. Conversely, 
$$v_1v_2 = \theta_1(w_2^*) \cdot \theta_1\theta_2(w_1^*) \cdot
\eps_{\theta_1,\theta_2}^\pm x_1\theta_1(x_2)v = v,$$
by the unit property of Q-systems. Hence $\D^\pm$ is generated by $\A$
and $v_1$ and $v_2$. The product $v_2v_1$ is given by 
$$v_2v_1 =  w_1^* \cdot \theta_1\theta_2(\theta_1(w_2^*)) \cdot
\eps_{\theta_1,\theta_2}^\pm x_1\theta_1(x_2)v =
\eps_{\theta_1,\theta_2}^\pm v.$$
Comparing these identities, \eref{comm-v} follows. 
\end{proof}

\begin{remark}
$A_1\times^+ A_2$ is unitarily equivalent to $A_2\times^- A_1$ by the unitary
$\eps_{\theta_1,\theta_2}$. 
\end{remark}

Rewriting \eref{comm-v} in terms of the 
charged field decomposition of $v_1$ and $v_2$, one has equivalently
(again suppressing possible multiplicity indices)
\be\label{comm-psi}
\psi_{2,\sig}\psi_{1,\rho}=\eps^{\pm}_{\rho,\sig}\cdot\psi_{1,\rho}
\psi_{2,\sig}. 
\ee
\begin{remark}
In view of \eref{transport}, these commutation relations are
compatible with the transport, i.e., if they hold for any $\rho$,
$\sig$, then they hold for every unitarily equivalent pair. 
In particular, as the statistics operators $\eps_{\rho,\sig}=\mathbf{1}$
whenever $\sig$ is localized to the left of $\rho$, left (right)
locality of $\psi_1$ w.r.t.\ $\psi_2$ implies \eref{comm-psi} with the
$-$ sign ($+$ sign) for arbitrary localization of $\rho$ and $\sig$. 
\end{remark}

Therefore, we have 

\begin{proposition} \label{p:product-ll-rl}
Within $\D^+=\B_1\times^+\B_2$, $\B_1$ is right-local w.r.t.\ $\B_2$, 
whereas within $\D^-=\B_1\times^-\B_2$, $\B_1$ is left-local w.r.t.\ $\B_2$. 
Every irreducible extension containing both $\B_1$ and $\B_2$ as intermediate 
extensions such that 
$\B_1$ is right-local (left-local) w.r.t.\ $\B_2$, is a quotient of 
$\D^+$ ($\D^-$).  
\end{proposition}

\begin{proof}
The first statement follows from \eref{comm-psi} and the
stated triviality of the statistics operator for charged fields with
the respective relative localization, along with the fact $\A$ is
local and both $\B_i$ are relatively local w.r.t.\ $\A$. The second
statement is due to the fact that \eref{comm-psi}, or equivalently 
\eref{comm-v} is the only independent relation among the generators 
$v_1$ and $v_2$ of $\D$, using the multiplication law of the product
Q-system. 
\end{proof}

Notice that even if both $\B_i$ are local, neither $\D^+$ nor $\D^-$
will be local in general. However, we know how to construct 
maximal local subtheories in $\D^\pm$ by the construction in the 
\sref{s:max}. Namely, one will have to determine the
centre(s) of the braided product Q-systems. The following partial
result is true without model specific knowledge.

\begin{lemma} \label{l:productQ-cent}
If $\B_2$ is local, then the right centre $(\D^+)\loc^-$ of 
$\D^+=\B_1\times^+\B_2$ contains at least $\B_2$. The same is true for
the left centre $(\D^-)\loc^+$ of $\D^-=\B_1\times^-\B_2$:
$$\B_2\subset (\D^+)\loc^-,\qquad \B_2\subset (\D^-)\loc^+.$$
Similarly, if $\B_1$ is local, then
$$\B_1\subset (\D^+)\loc^+,\qquad \B_1\subset (\D^-)\loc^-.$$
\end{lemma}
Similar statements are in general not true for the other combinations,
e.g., $\B_2$ local will not belong to the left centre $(\D^+)\loc^+$ (=
relative commutant of left wedges) of the $\times^+$ braided product,
or to $(\D^-)\loc^-$. See, however, the \sref{s:alpha} for 
the local subtheories $(\D^+)\loc^+$ when $\B_1$ is chiral and 
$\B_2$ is the canonical local extension.

\begin{proof}
This can be seen by inspection of Fig.\ 1: an element of
$\B_2$ in the double cone commutes with every element of $\B_2$ in the
right wedge because $\B_2$ is local, and with every element of $\B_1$ in the
right wedge by \pref{p:product-ll-rl}.
\end{proof}
$$\tikzmatht{          
          \fill[black!10] (-5,-3.5) rectangle (3,3.5);
          \draw[thick] (-1,0) -- (-2.5,1.5) -- (-4,0) -- (-2.5,-1.5)
          --(-1,0);
          \fill[black!15] (3,3.5) -- (2.5,3.5) -- (-1,0) -- (2.5,-3.5)
          -- (3,-3.5);
          \draw[thin] (-1,0) -- (2.5,-3.5); \draw[thin] (-1,0) -- (2.5,3.5);
          \draw[thin] (-2.5,1.5) -- (-.5,3.5);           
          \draw[thin] (-2.5,-1.5) -- (-.5,-3.5); 
          \draw[thick] (-2,.6) -- (-2.5,1.1) -- (-3,.6) -- (-2.5,.1)
          --(-2,.6);            
          \draw[thick] (-2,-.6) -- (-2.5,-1.1) -- (-3,-.6) -- (-2.5,-.1)
          --(-2,-.6);            
\node at (-2.5,-.6){\tiny 2}; \node at (-2.5,.6){\tiny 1};
\node at (2.2,0){\small$\times^+$};
          \draw[thick] (2,.6) -- (1.5,1.1) -- (1,.6) -- (1.5,.1)
          --(2,.6);          
          \draw[thick] (2,-.6) -- (1.5,-1.1) -- (1,-.6) -- (1.5,-.1)
          --(2,-.6);          
\node at (1.5,-.6){\tiny 2}; \node at (1.5,.6){\tiny 1};
\node at (-1.8,0){\small$\times^+$};
}
\supset
\tikzmatht{          
          \fill[black!10] (0,-3.5) rectangle (5,3.5);
          \draw[thick] (1,0) -- (2.5,1.5) -- (4,0) -- (2.5,-1.5)
          --(1,0);
          \draw[thick] (2,-.6) -- (2.5,-1.1) -- (3,-.6) -- (2.5,-.1)
          --(2,-.6);            
\node at (2.5,-.6){\tiny 2};
\node at (3.9,-3) {\bf Fig.\ 1};}
$$
{\bf Fig.\ 1}: 
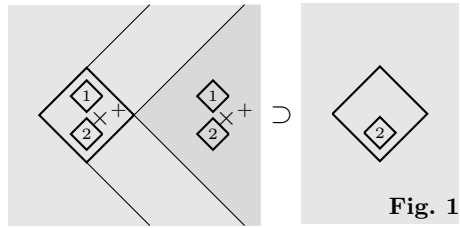
This figure visualizes the statement of \lref{l:productQ-cent}: If
$\B_2$ is local, then the right centre (computed by relative commutants 
of right wedge algebras) of the braided product of $\B_1$ and $\B_2$ 
contains $\B_2$ if and only if $\B_2$ is left-local w.r.t.\ $\B_1$ (i.e., 
the braided product is the $\times^+$ product by \pref{p:product-ll-rl}).

\begin{lemma} \label{l:productQ-cent2}
Let $\A_{\rm 2D}=\A\otimes\A$, where $\DHR(\A)$ is modular. If $A_1$ is a
chiral Q-system of the form $A\otimes 1$, i.e., $\B_1$ is a chiral
extension $\A_{\rm 2D}\subset \B\otimes\A$, and $A_2$ is the canonical
Q-system $R$ of \pref{p:canonical}, i.e., $\A_{\rm 2D}\subset \B_2$ is the canonical extension, then one has
equality in \lref{l:productQ-cent}: 
$$(\D^+)\loc^- = (\D^-)\loc^+ = \B_2.$$
\end{lemma}

\begin{proof}
One can compute the trace (cf.\ \sref{s:Q-ext})
$\Tr(p^\pm) \equiv R^*p^\pm R = R^*\Theta(p^\pm)R$ of the centre projection. 
By using the fact that the braiding is non-degenerate, i.e., 
$\eps_{\sig,\rho}\eps_{\rho,\sig}=\mathbf{1}$ for all $\rho$ implies
$\sig=\id$, one computes $\Tr(p^\pm) = d_R^2 =\dim(R)$ for the left
centre of the $\times^-$ product, and for the right centre of the
$\times^+$ product. Therefore the canonical extension exhausts the
centre projection.
\end{proof}

\subsection{The $\alpha$-induction construction and the full centre}
\label{s:alpha}

A special case of the centre is the {\bf full centre} \cite[Def.\ 2.6]{KR}.
If the braided category has only finitely many irreducible 
objects, then the canonical Q-system $R$ (\pref{p:canonical}) 
is a commutative Q-system in $\C\boxtimes\C\opp$. Every Q-system $A$ in 
$\C$ lifts trivially to a Q-system in $\C\boxtimes\C\opp$ by tensoring 
with the identity as in \lref{l:productQ-cent2}. Then the (left) full centre of $A$ is defined
\cite{FFRS06} as 
\be\label{fullc}
Z^+[A]=C^+[(A\otimes 1)\times^+R],\ee
i.e., the left centre of the $\times^+$-braided product of the lifted 
Q-system $A$ with the canonical Q-system. The right full centre,  
defined analogously as $Z^-[A]=C^-[(A\otimes 1)\times^-R]$,
is not independent: one has $Z^-[A]=Z^+[\widehat A]$
where $\widehat A$ is the Q-system $(\theta,w,\eps_{\theta,\theta}\,x)$.
In the sequel, ``full centre'' $Z[A]$ will always mean the left full centre.

The full centre has many remarkable properties as a mathematical
object of its own. In particular, it is a complete Morita invariant:
if $\C$ is modular, then two Q-systems have the same full centre if
and only if they are Morita equivalent \cite[Thm.\ 1.1]{KR}. 

Here, we shall be interested only in its relevance for quantum field
theory. (Recall that $\DHR(\A)$ is modular \cite{KLM} for every
completely rational theory, hence the full centre is indeed a Morita
invariant.) The full centre of a Q-system $A$ in the DHR category of 
a chiral net $\A$ is a Q-system in $\DHR(\A\otimes\A)$, hence it 
describes a local extension of $\A_{\rm 2D}=\A\otimes \A$. By considering the 
algebraic interpretation \pref{p:productQ} and
\pref{p:relcomm=cent} of the two steps: braided product and left
centre, involved in the definition \eref{fullc}, the full centre
extension is obtained by the corresponding steps: 
\begin{enumerate}\itemsep0mm
\item[(i)] Construction of a 2D nonlocal CFT as a braided product of 
a chiral extension with the canonical 2D local QFT, and 
\item[(ii)] restriction to a maximal local subtheory by taking relative 
commutants of the algebras associated with nested wedge regions.  
\end{enumerate}
It is therefore an instance of the combinations ``(+,+)'' and
``$(-,-)$'' {\em not} covered by \lref{l:productQ-cent}. 

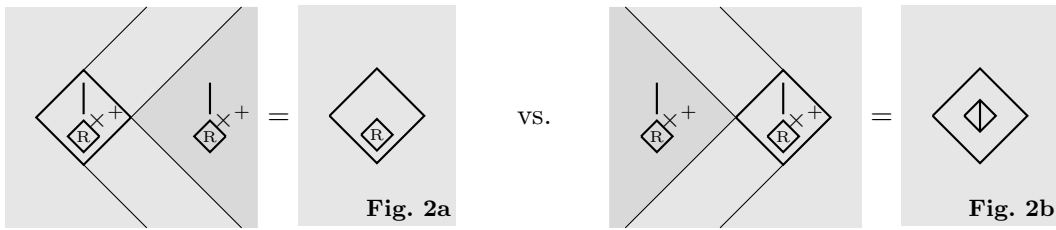
$$\tikzmatht{          
          \fill[black!10] (-5,-3.5) rectangle (3,3.5);
          \draw[thick] (-1,0) -- (-2.5,1.5) -- (-4,0) -- (-2.5,-1.5)
          --(-1,0);
          \fill[black!15] (3,3.5) -- (2.5,3.5) -- (-1,0) -- (2.5,-3.5)
          -- (3,-3.5);
          \draw[thin] (-1,0) -- (2.5,-3.5); \draw[thin] (-1,0) -- (2.5,3.5);
          \draw[thin] (-2.5,1.5) -- (-.5,3.5); 
          \draw[thin] (-2.5,-1.5) -- (-.5,-3.5); 
          \draw[thick] (-2.5,.1) -- (-2.5,1.1);            
          \draw[thick] (-2,-.6) -- (-2.5,-1.1) -- (-3,-.6) -- (-2.5,-.1)
          --(-2,-.6);            
          \node at (-2.5,-.6){\tiny R};\node at (-1.8,0){\small$\times^+$};
          \draw[thick] (1.5,1.1) -- (1.5,.1);          
          \draw[thick] (2,-.6) -- (1.5,-1.1) -- (1,-.6) -- (1.5,-.1)
          --(2,-.6);          
          \node at (1.5,-.6){\tiny R};\node at (2.2,0){\small$\times^+$};
}
= 
\tikzmatht{          
          \fill[black!10] (0,-3.5) rectangle (5,3.5);
          \draw[thick] (1,0) -- (2.5,1.5) -- (4,0) -- (2.5,-1.5)
          --(1,0);
          \draw[thick] (2,-.6) -- (2.5,-1.1) -- (3,-.6) -- (2.5,-.1)
          --(2,-.6);            
\node at (2.5,-.6){\tiny R};
\node at (3.5,-3) {\bf Fig.\ 2a};
}
\qquad\hbox{vs.}\qquad
\tikzmatht{          
          \fill[black!10] (-3,-3.5) rectangle (5,3.5);
          \draw[thick] (1,0) -- (2.5,1.5) -- (4,0) -- (2.5,-1.5)
          --(1,0);
          \fill[black!15] (-3,3.5) -- (-2.5,3.5) -- (1,0) --
          (-2.5,-3.5) -- (-3,-3.5);
          \draw[thin] (1,0) -- (-2.5,-3.5); \draw[thin] (1,0) -- (-2.5,3.5);
          \draw[thin] (2.5,1.5) -- (.5,3.5); \draw[thin] (2.5,-1.5) --
          (.5,-3.5); 
          \draw[thick] (2.5,.1) -- (2.5,1.1);            
          \draw[thick] (2,-.6) -- (2.5,-1.1) -- (3,-.6) -- (2.5,-.1)
          --(2,-.6);            
\node at (2.5,-.6){\tiny R};\node at (3.2,0){\small$\times^+$};
          \draw[thick] (-1.5,1.1) -- (-1.5,.1);          
          \draw[thick] (-2,-.6) -- (-1.5,-1.1) -- (-1,-.6) -- (-1.5,-.1)
          --(-2,-.6);          
\node at (-1.5,-.6){\tiny R};\node at (-.8,0){\small$\times^+$};
}
= 
\tikzmatht{          
          \fill[black!10] (0,-3.5) rectangle (5,3.5);
          \draw[thick] (1,0) -- (2.5,1.5) -- (4,0) -- (2.5,-1.5)
          --(1,0);
          \draw[thick] (2,0) -- (2.5,.5) -- (3,0) -- (2.5,-.5)
          --(2,0);            
          \draw[thick] (2.5,-.5) -- (2.5,.5); 
\node at (3.5,-3) {\bf Fig.\ 2b};
}
$$
{\bf Fig.\ 2a:} Same as Fig.\ 1, with $\B_1$ some chiral extension,
and $\B_2$ the canonical local extension. By nondegeneracy of the
braiding, the right centre of the braided product (= relative
commutant of right wedges) is exactly the canonical extension 
(\lref{l:productQ-cent2}). \\ 
{\bf Fig.\ 2b:} The full centre of the chiral theory is the left centre 
(= relative commutant of left wedges) of the
braided product with the canonical extension chiral extension. It
equals the $\alpha$-induction construction from the chiral extension. 
It is in general different from, but may happen to be isomorphic to the 
canonical local extension. Being contained in the braided product, it 
is generated by certain products of canonical and chiral fields.

$$\tikzmatht{
          \fill[black!10] (-5,-5) rectangle (10,5);
          \fill[black!15] (-4,2) rectangle (-1.2,3);
          \draw[thick] (-4,2) rectangle node {chiral} (-1.2,3);
          \node at (-.5,2.5){$\times^+$};
          \fill[black!15] (0.2,2) rectangle (4,3);
          \draw[thick] (0.2,2) rectangle node {$R$, local} (4,3);
          \node at (4.5,2.5){$=$};
          \fill[black!15] (5,2) rectangle (8.5,3);
          \draw[thick] (5,2) rectangle node {half-local} (8.5,3);
          \draw[thick] (7,2)--(5,-2);  \draw[thick] (7,2)--(8,-2);
          \fill[black!15] (-.5,0) rectangle (8.5,-1);
          \draw[thin] (-.5,0) rectangle node {maximal local subtheories}
          (8.5,-1);
          \fill[black!15] (0,-2) rectangle (6,-3);
          \draw[thick] (0,-2) rectangle node {``full centre''} (6,-3);
          \fill[black!15] (7.5,-2) rectangle (8.5,-3);
          \draw[thick] (7.5,-2) rectangle node {$R$} (8.5,-3);
          \node at (6,1){$p^+$};
          \node at (8,1){$p^-$};
\node at (8.5,-4.5) {\bf Fig.\ 3};}$$
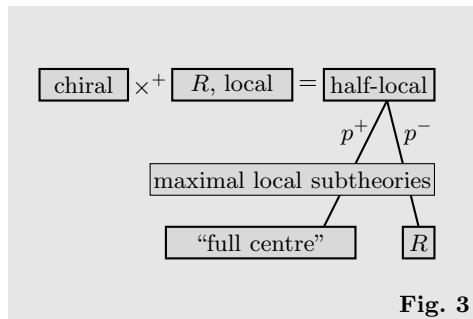
{\bf Fig.\ 3}: Another view of Fig.\ 2.

It was recently recognized \cite{BKL} that the full centre of the 
Q-system for the chiral extension $\A\subset\B$ coincides with the 
``$\alpha$-induction construction''. 

The latter is a generalization of the canonical Q-system
(\pref{p:canonical}) for the construction of two-dimensional (2D)  
extensions \cite{R00}. It was found by solving the defining identities of 
a Q-system in $\DHR(\A\otimes \A)$ with the help of 
$\alpha$-induction. $\alpha$-induction is a prescription to extend 
DHR endomorphisms $\rho$ of the underlying chiral theory $\A$ to 
endomorphisms $\alpha^\pm_\rho$ of a chiral extension $\B$ of $\A$. 
The latter depend on the braiding $\eps^\pm$, and are in general not DHR 
endomorphisms. From these data, one obtains Q-systems with coupling matrix 
(\eref{Z}) $Z_{\sig,\ol \tau}=\dim\Hom(\alpha^+_\sig,\alpha^-_\tau)$. If $\A$
is completely rational, then $Z$ is a modular invariant matrix
\cite{BEK}. 

This ``arithmetic'' construction may be regarded as not very
satisfactory from the AQFT point of view. However, combining the 
above algebraic interpretation of the full centre, as well as the 
results in \cite{BKL} and \cite{LR09}, we conclude:

\begin{corollary} The following constructions of two-dimensional local
  extensions of $\A_{\rm 2D}=\A\otimes\A$, where $\A$ is completely rational, are all equivalent: 
\begin{enumerate}\itemsep0mm
\item[(1)] The extension corresponding to the ``full centre'' of a chiral Q-system.
\item[(2)] The ``$\alpha$-induction construction'' of a 2D local CFT associated
  with a (possibly nonlocal) chiral extension \cite{R00}.
\item[(3)] The construction of a 2D local CFT by ``removing the
  boundary'' via a limit state on a (hard) boundary CFT \cite{LR09}.
\item[(4)] The construction of a 2D local CFT by the two steps (i) and
  (ii) above. 
  \end{enumerate}
\end{corollary}

\begin{proof}
The equivalence (1) $\LRA$ (2) is established in \cite{BKL}, 
by identification of the Q-systems. The equivalence (2) $\LRA$
(3) was proven in \cite{LR09}. The equivalence (1) $\LRA$ (4) is an
immediate consequence of the definition of the full centre and the
algebraic interpretation of the two operations involved, as discussed
before.  
\end{proof}

Now, (4) provides a more satisfactory direct algebraic understanding 
(and in fact, a surprisingly simple one) of the previous indirect 
constructions (2) and (3) of two-dimensional extensions with 
modular-invariant coupling matrices \eref{Z}. 

Moreover, the construction (3) together with the maximality result for hard
boundaries in \cite{LR04} implies that every maximal irreducible local
extension is of this form.

Let $\A\subset \B$ be a chiral (non-local) extension and 
$\A\otimes\A\subset \B_{\rm 2D}$ the 2D local CFT obtained by its full centre.
Replacing $R$ by the trivial Q-system in \eref{fullc}, we can conclude that
$C^+[A]\otimes 1$ is contained in the full centre $Z^+[A]$, i.e., there is an intermediate extension $\A\otimes \A \subset
\B\loc^+\otimes \A\subset \B_{\rm 2D}$. Similarly, doing the same with  
the $(-,-)$ case, we get $\A\otimes \A \subset \A\otimes
\B\loc^-\subset \B_{\rm 2D}$. Together we recover the well-known
tensor product of maximal chiral extensions: 
\be \label{eq:MaximalChiral}
\A \otimes\A\subset \B\loc^+ \otimes \B\loc^-\subset \B_{\rm 2D}.
\ee

\subsection{The relative commutant of product extensions}
\label{s:decoP}

The braided product $A_1\times^\pm A_2$ of two irreducible Q-systems 
($\dim\Hom(\id,\theta_i)=1$) will in general be reducible, because 
$\dim\Hom(\id,\theta_1\theta_2) = \dim\Hom(\theta_1,\theta_2)>1$, 
cf.\ \sref{s:decoQ}. For the corresponding extensions, this
means $\A'\cap\D^\pm\neq \CC\cdot\mathbf{1}$ because, for 
$\rho\prec\theta_1$ and $\rho\prec\theta_2$, operators of the 
form $\psi_{1,\rho}^*\psi_{2,\rho}$ commute with every $a\in\A$ by \eref{charged}. 

\begin{proposition} \label{p:prod-loc}
The relative commutant $\A'\cap \D^\pm$ of the braided product extension
$\D^\pm$ is spanned by all operators of the form  
$\psi_{1,\rho}^{n*}\psi_{2,\rho}^m$, where $\rho\prec\theta_1$,
$\rho\prec\theta_2$, and $n=1,\dots\dim\Hom(\rho,\theta_1)$,
$m=1,\dots\dim\Hom(\rho,\theta_2)$. If both Q-systems 
$A_1$ and $A_2$ are commutative (hence the corresponding extensions $\B_i$
are local), then the relative commutant $\A'\cap \D^\pm$ coincides with 
the centre\footnote{The centre here is the algebraic centre of a C*
  algebra, not to be confused with the centre of a Q-system!}  
$\D^\pm{}'\cap \D^\pm$. 
\end{proposition}

\begin{proof} 
In \lref{l:center}, we have characterized the relative commutant by
the elements $\iota(q)v$ with $q\in\Hom(\theta,\id)$, where 
$\theta=\theta_1\theta_2$ for a product Q-system. By Frobenius
reciprocity, we have
$\Hom(\theta_1\theta_2,\id)=r_1^*\theta_1(\Hom(\theta_2,\theta_1))$. A 
basis of $\Hom(\theta_2,\theta_1)$ is given by $w_{1,\rho}w_{2,\rho}^*$
with $w_{i,\rho}\in\Hom(\rho,\theta_i)$. With $\psi_{1,\rho} =
w_{1,\rho}^*\theta_1(w_2)^*v^\pm$ and $\psi_{2,\rho} =
w_{1}^*\theta_1(w_{2,\rho}^*)v^\pm$ (suppressing the symbols $\iota$
for simplicity), one computes  
$$\psi_{1,\rho}^*\psi_{2,\rho} = v^\pm{}^*\theta_1(w_2)w_{1,\rho}
w_{1}^*\theta_1(w_{2,\rho}^*)v^\pm =
r^*\theta_1\theta_2\big(\theta_1(w_2)w_{1,\rho} 
w_{1}^*\theta_1(w_{2,\rho}^*)\big)x^\pm v^\pm = qv^\pm$$
with $q=r_1^*\theta_1(w_{1,\rho}w_{2,\rho}^*)$. Since $q\in
\Hom(\theta_1\theta_2,id)$ arises by  
Frobenius reciprocity from a basis $w_{1,\rho}w_{2,\rho}^*$ of
$\Hom(\theta_1,\theta_2)$,  
the operators $\psi_{1,\rho}^*\psi_{2,\rho}$ form a basis of
$\Hom(\theta_1\theta_2,\id)v = \A'\cap \D$.  

In \lref{l:center}, we have characterized the centre by
$q\in\Hom(\theta,\id)$ satisfying  
the supplementary relation \eref{centre-char}. 
For $q\in\Hom(\theta_1\theta_2,\id)$ one computes 
$$q x^\pm = \theta_1(q)\eps_{\theta_1,\theta_1}^\mp x_1\theta_1(x_2) =
\theta_1(q)x_1\theta_1(x_2) = 
\theta_1(q)x_1\theta_1(\eps_{\theta_2,\theta_2}^\mp x_2) = 
\theta_1\theta_2(q) x^\pm $$
if both Q-systems are commutative, using \eref{local}. 
Thus the supplementary condition \eref{centre-char} for the product 
Q-system is satisfied by every $q\in\Hom(\theta,\id)$. 
\end{proof} 

This result will become important in the treatment of phase boundaries 
(cf.\ \sref{s:Bound}).

Every minimal projection $e$ in the centre of $\D$ defines an
irreducible representation $\pi_e(\D)=e\D$ of $\D$, and hence also of
the two intermediate local extensions $\A\subset\B_1$  and
$\A\subset\B_2$. 

\section{Phase boundaries}
\label{s:Bound} \setcounter{equation}{0}

As discussed in \sref{s:phaseb}, causality at a phase boundary
requires only that the local QFT to the left of the boundary is
left-local w.r.t.\ the local QFT to the right of the boundary. Such
situations are precisely constructed by the braided products of extensions. 

This is in contrast to previous constructions, like \sref{s:alpha},
where braided product extensions were used as 
intermediate nonlocal constructions from which one has to descend 
to the left or right centre in order to get (globally) local subtheories. 

\subsection{The universal construction}
\label{s:universal}

A phase boundary is a transmissive boundary with chiral observables
$\A_{\rm 2D}=\A_+\otimes\A_-$. The phases on both sides of the boundary are given by
a pair of Q-systems $A^L =(\Theta^L,W^L,X^L)$ and
$A^R=(\Theta^R,W^R,X^R)$ in $\DHR_0(\A_{\rm 2D})$, describing local 2D
extensions $\A_{\rm 2D}\subset \B_{\rm 2D}^L$ and $\A_{\rm 2D}\subset \B_{\rm 2D}^R$.

Now consider the braided product Q-systems (cf.\ \sref{s:product})
\be\label{product}
(\Theta=\Theta^L\circ\Theta^R,W=W^L\times
W^R,X=(1\times\eps^\pm_{\Theta^L,\Theta^R}\times 1)\circ(X^L\times X^R))
\ee
and the corresponding extensions $\A_{\rm 2D}\subset\D_{\rm 2D}^\pm$.
The original extensions $\B_{\rm 2D}^L$, $\B_{\rm 2D}^R$ are intermediate, cf.\
\sref{s:product}:
\be\label{boundalg}
\A_{\rm 2D}\subset \B_{\rm 2D}^L\subset \D_{\rm 2D}^\pm\,\qquad\A_{\rm 2D}\subset \B_{\rm 2D}^R\subset
\D_{\rm 2D}^\pm,
\ee
and the nets $\D_{\rm 2D}^\pm$ are generated by $\A_{\rm 2D}$ and two sets of
charged fields $\Psi^L_{\sig\otimes\tau}$ ($\sig\otimes\tau\prec\Theta^L$)
and $\Psi^R_{\sig\otimes\tau}$ ($\sig\otimes\tau\prec\Theta^R$),
suppressing possible multiplicity indices. The braided product
Q-system determines their commutation relations among each other as in
\eref{comm-psi}: 
\be\label{CR}
\Psi^R_{\sig\otimes\tau}\Psi^L_{\sig'\otimes\tau'} =
\eps^\pm_{\sig'\otimes\tau',\sig\otimes\tau}\cdot
\Psi^L_{\sig'\otimes\tau'}\Psi^R_{\sig\otimes\tau}.
\ee

By \eref{trivbraid2}, $\eps^-_{\sig'\otimes\tau',\sig\otimes\tau}=\mathbf{1}$
whenever $\sig'\otimes\tau'$ is localized to the spacelike left of  
$\sig\otimes\tau$. Thus, the choice of $\eps^-$ in \eref{CR} ensures
that $\B^L$ is left-local w.r.t.\ $\B^R$, as required by causality. In
accord with \pref{p:product-ll-rl}, this dictates our choice of
the braided product Q-system to be 
\be\label{phaseproduct}
(\Theta^L,W^L,X^L)\times^-(\Theta^R,W^R,X^R),
\ee
or -- equivalently -- $(\Theta^R,W^R,X^R)\times^+(\Theta^L,W^L,X^L)$.

The conclusion is our main structural result: 

\begin{proposition} \label{p:universal} (``Universal construction'')
The extension $\D$ of $\A$ defined by the Q-system \eref{phaseproduct}
implements a boundary condition in the sense of
\dref{d:bc}. It is universal in the sense that every irreducible 
boundary condition appears as a representation of $\D$. 
\end{proposition}

\begin{proof} By \pref{p:productQ}, $\D$ is generated by $\B^L$
and $\B^R$, and by \pref{p:product-ll-rl}, $\B^L$ is left-local
w.r.t.\ $\B^R$. The universal property follows because apart from the
identification of the subalgebras $\A\subset\B^L$ and $\A\subset\B^R$,
the commutation relation \eref{comm-v}, i.e., left locality, is the only
additional relation among the generators of $\D$ imposed by the
product Q-system, cf.\ \pref{p:productQ}.
\end{proof}

\subsection{Boundary conditions}
\label{s:bcond}

The treatment of boundary conditions as defined in \dref{d:bc} in this 
subsection is quite general. It applies  
to boundaries between two-dimensional conformal QFTs described by
local extensions $\B_{\rm 2D}^L$ and $\B_{\rm 2D}^R$ of $\A=\A_+\otimes\A_-$, as
well as to ``boundaries'' between two chiral QFTs, i.e., two
chiral QFTs $\B^L$ and $\B^R$ which are local chiral extensions of an 
underlying local chiral net $\A$, separated by a point. 
\pref{p:universal} litterally applies also as a universal construction
of boundary conditions for the latter case.

In \sref{s:classi}, we shall present a more powerful result, 
\tref{t:center}, which only pertains to two dimensions. Namely, 
it covers the case when the underlying net is $\A\otimes\A$ and 
$\B^L_{\rm 2D}$ and $\B^R_{\rm 2D}$ are maximal two-dimensional local extensions, 
namely $A^L$ and $A^R$ both are full centres of chiral Q-systems.   

For the remainder of {\em this} section, we drop the 
subscript distinguishing two-dimensional nets.

Let $\A\subset \B^L$ and $\A\subset \B^R$ be the local
extensions of $\A$ defining the QFT to the left and to the right of
the boundary. Let $\D\equiv \D^-$ be the universal construction, namely the
(nonlocal) extension obtained by the braided product $A^L\times^- A^R$
of the corresponding commutative Q-systems $A^L$ and $A^R$.  

By \pref{p:prod-loc}, the extension $\A\subset \D$ is in general
reducible, and the relative commutant coincides with the centre
$\D'\cap \D$. The central decomposition gives irreducible representations 
of $\D$, restricting to representations of both $\B^L$ and $\B^R$ on a
common Hilbert space. 

Let $\Theta^L=\bigoplus_{\rho} Z^L_{\rho}\cdot\rho$ and 
$\Theta^R=\bigoplus_{\rho} Z^R_{\rho}\cdot\rho$ be the irreducible
decomposition. (In the two-dimensional case, each $\rho$ is of the form
$\rho^+\otimes\rho^-$.) By \pref{p:prod-loc}, the dimension of the centre is
$\dim\Hom(\Theta^L,\Theta^R) = \sum_\rho Z^L_{\rho}\cdot Z^R_{\rho}$,
and there are as many irreducible representations, implementing
different boundary conditions. 
 
\begin{lemma} \label{l:mult1} Each minimal projection $e\in\D'\cap \D$ defines
  an irreducible extension $\A=\A e\subset \D e$. In
  particular, the representation $\D e$ contains a unique vacuum vector.
\end{lemma}

\begin{proof}
Since every reduced Q-system contains the identity {\em at least} with
multiplicity one, and the number of reduced Q-systems equals the
multiplicity of the identity in $\Theta=\Theta^L\Theta^R$, it follows that 
$\dim\Hom(\id,\Theta_e)=1$ for each minimal projection. Therefore, the
extension $\A e\subset \D e$ is irreducible, and the
vacuum representation of $\A$ is contained {\em exactly} with
multiplicity one.
\end{proof}

Together with the universality property, \pref{p:universal}, we
conclude 
\begin{corollary} \label{c:bc=mp}
The minimal central projections of the braided
  product between two local QFTs classify the irreducible 
  boundary conditions.  
\end{corollary}

From \pref{p:prod-loc}, we know the centre as a linear space. A
basis of this space (see below) are the ``neutral'' products of
charged field operators of the form 
\be\label{Brho} B_\rho = \Psi_{\rho}^{L*}\Psi_{\rho}^R,
\ee 
where
$\rho\prec\Theta^L$ and $\rho\prec\Theta^R$. These
operators commute with $\iota(\A)$ and hence belong to the centre by 
\pref{p:prod-loc}. In an irreducible representation, they are
multiples of $\mathbf{1}$. Thus, every representation comes with a
characteristic set of relations  
between the charged fields  $\Psi_\rho^{L}$ and $\Psi_\rho^R$. These
relations are the desired boundary conditions. 

In order to determine the values of the central operators $B_\rho$ in
each representation, we need to compute $B_\rho$ as linear
combinations of the minimal central projections $E_m$ in $\D$:
\be\label{piF}
B_\rho = \sum_m \pi_m(B_\rho) \cdot E_m,
\ee
with $\pi_m(B_\rho) \in\CC$. This meaning of the expansion
coefficients follows because $E_m$ is represented by $\mathbf{1}$ in the
representation $\pi_m$, and by $0$ in all other representations. 

In order to determine the minimal central projections $E_m$, we need 
to know the centre as an algebra. At this point, a reformulation of
the task is convenient.

We introduce the linear bijection 
\be\label{chi}
\chi:\Hom(\Theta^R,\Theta^L)\to \D'\cap \D,\qquad T \mapsto
\iota(R^{L*}\Theta^L(T))\cdot V.
\ee
Then one has 
\be\label{chiprod}
\chi(T_1)\cdot\chi(T_2) = \chi(T_1\ast T_2)
\ee
where the associative $\ast$-product in $\Hom(\Theta^R,\Theta^L)$ is 
defined as 
\be\label{ast}
T_1\ast T_2 := X^{L*}\cdot \Theta^L(T_1)T_2\cdot X^R \equiv
X^{L*}\cdot T_2\Theta^R(T_1)\cdot X^R.
\ee
This associative product is also commutative, because the Q-systems $A^L$ and
$A^R$ are commutative. The unit of the $\ast$-product is
$W^{L}W^{R*}$. Furthermore,
\be\label{chi*}
\chi(T)^* = \chi(F(T^*))
\ee
where $F:\Hom(\Theta^L,\Theta^R)\to\Hom(\Theta^R,\Theta^L)$
is the Frobenius conjugation $T^*\mapsto R^{R*}\Theta^R(T^*R^L) =
\Theta^L(R^{R*}T^*)R^L$. 

Thus, the task is to find $I_m=\chi\inv(E_m)$ which are the minimal
projections in $\Hom(\Theta^R,\Theta^L)$ with respect to the
$\ast$-product, and expand $\chi\inv(B_\rho)$ in terms of $I_m$. 

\medskip

Let $\Psi_\rho^L= W_\rho^{L*}\Theta^L(W^{R*})V$, $\Psi_\rho^R=
W_\rho^{R*}W^LV$ with $W^Y_{\rho}\in\Hom(\rho,\Theta^Y)$ ($Y=L,R$), as
in \sref{s:charged}. We suppress possible multiplicity indices
if these spaces are more than one-dimensonal, and assume the charged
field operators to be isometries, i.e., $W^Y_{\rho}$ are normalized as
$W^Y_{\rho}{}^*W^Y_{\rho}=d_Y/\dim(\rho)$ in accordance with
\eref{wnorm}. $d_Y=\sqrt{\dim(\Theta^Y)}$ are the dimensions of the
factor Q-systems.  
 
By Frobenius reciprocity
$W^Y_{\ol\rho}:=R^*_\rho\rho(W^{Y*}_{\rho}R^Y)
=\Theta^Y(R^*_\rho W^{Y*})R^Y\in\Hom(\rho,\Theta^Y)$ 
with the same normalization, and $\Psi^Y_{\ol\rho}$ and $B_{\ol\rho}$
are the corresponding charged fields and central operators. 

In particular, \eref{Brho} turns into 
\be\label{chiF}
B_\rho\equiv \Psi_{\rho}^{L*}\Psi_{\rho}^R =\chi(W^{L}_\rho W_\rho^{R*}).
\ee
Since $W^{L}_\rho W_\rho^{R*}$ span $\Hom(\Theta^R,\Theta^L)$,
$B_\rho$ span $\D'\cap \D$. 

\medskip

The following proposition gives the algebraic relations among $B_\rho$
in terms of the pair of Q-systems $A^L$, $A^R$. Knowing their algebra,
one can compute their decomposition into central projections. 

\begin{proposition} \label{p:center=psipsi} The algebraic relations of
  the centre $\D'\cap \D$ of the braided product of two local
  extensions (corresponding to two commutative Q-systems) are
\be\label{centerF}
B_\rho^*=B_{\ol\rho},\qquad B_\rho\cdot B_\sig = \sum_{\tau\prec\rho\sig}
f_{\rho,\sig}^\tau \cdot B_\tau
\ee
where 
$$f_{\rho,\sig}^\tau = f_{\sig,\rho}^\tau =
\frac{\dim(\tau)^2}{d_Ld_R}\cdot W_\tau^{L*}\cdot 
\Big((W^L_{\rho}W^{R*}_{\rho})\ast(W^L_{\sig}W^{R*}_{\sig})\Big)\cdot
W_\tau^R.$$
\end{proposition}

The expressions given for $f_{\rho,\sig}^\tau$ are self-intertwiners
$\in\Hom(\tau,\tau)$, hence multiples of $\mathbf{1}$.

\begin{proof} 
We give only a sketch of the proof. The first relation
follows, in view of \eref{chi*}, from $F(W^R_{\rho}W^{L*}_{\rho}) = 
W^L_{\ol\rho}W^{R*}_{\ol\rho}$ . The second relation is just the 
expansion of $(W^L_{\rho}W^{R*}_{\rho})\ast(W^L_{\sig} W^{R*}_{\sig})$ 
in terms of $W^L_{\tau}W^{R*}_{\tau}$, by using the stated 
normalizations of $W_\rho^Y$, translated in view of \eref{chiF}.  
\end{proof}

Up to normalization factors, the coefficients $f_{\rho,\sig}^\tau$ are
just products of the expansion coefficients $\zeta^R$ and
$\ol{\zeta^L}$ of the Q-systems, as in in \cite{LR95} or
\cite{R00}. The possible multiplicity indices are
  easily restored in \pref{p:center=psipsi}, just remembering that each
  $B_\rho$ becomes a tensor.

In the two-dimensional case, if $\A_+=\A_-$ is completely rational and
both Q-systems in the 
proposition equal the canonical Q-system in the category
$\DHR(\A)\boxtimes\DHR(\A)\opp$, each sector is of the form
$\rho\otimes\ol\rho$, there are no multiplicities, and  
the coefficients simplify drastically:
\be
f_{\rho\otimes\ol\rho,\sig\otimes\ol\sig}^{\tau\otimes\ol\tau}=\frac
  {\dim(\tau)}{\dim(\rho)\dim(\sig)} \cdot N_{\rho\sig}^\tau.
\ee
Thus the central operators $\dim(\rho) B_{\rho\otimes\ol\rho}$
represent the fusion rules. This implies that the various irreducible 
representations are labelled by the irreducible chiral sectors 
$\sig\in\DHR(\A)$, and \eref{piF} becomes
\be\label{verlinde}
\pi_\sig(\Psi^{L*}_{\rho\otimes\ol\rho}\Psi^R_{\rho\otimes\ol\rho})
= \frac{S_{\rho,\sig}S_{0,0}}{S_{\rho,0}S_{0,\sig}}\ee
(where $S_{0,\sig}=S_{\sig,0}=\dim(\sig)/d_R$), thus giving certain
angles between the isometric intertwiners $\Psi^L$ and $\Psi^R$. 
In particular, $\sig=\id$ gives
$\Psi^{L*}_{\rho\otimes\ol\rho}\Psi^R_{\rho\otimes\ol\rho}=\mathbf{1}$, hence
$\Psi^{L}_{\rho\otimes\ol\rho}=\Psi^R_{\rho\otimes\ol\rho}$, i.e.,
the trivial boundary where the left and right fields coincide.

\subsection{Example 5: Phase boundaries of the Ising model.}
\label{E5}
In the Ising model ($\Theta^L=\Theta^R=\Theta\can$) 
everything can be computed explicitly: $\D_{\rm 2D}$ is generated by $\A_{\rm 2D}$
and $\Psit^Y$ and $\Psis^Y$ ($Y=L,R$)
satisfying the relations as in Example 4, \sref{E4} for both $Y=L,R$ and mutual
commutation relations 
$$\Psit^R\Psit^L=\Psit^L\Psit^R,\quad\Psit^R\Psis^L=(u\otimes
u)\Psis^L\Psit^R,\quad\Psis^R\Psit^L=(u\otimes u)\Psit^L\Psis^R,$$
$$\Psis^R\Psis^L=((rr^*-itt^*)\otimes(rr^*+itt^*))\Psis^L\Psis^R.$$ 
The centre of $\D_{\rm 2D}$ is spanned by $\mathbf{1}$ and
$B_{\tau\otimes\tau}=\Psit^{L*}\Psit^R$ and
$B_{\sig\otimes\sig}=\Psis^{L*}\Psis^R$, satisfying
$B_{\tau\otimes\tau}^2=\mathbf{1}$, $B_{\tau\otimes\tau}
B_{\sig\otimes\sig}=B_{\sig\otimes\sig}$, and 
$B_{\sig\otimes\sig}^2=\frac12(1+B_{\tau\otimes\tau})$. 

Therefore, the minimal central projections are
$\frac12(1+B_{\tau\otimes\tau})\frac12(1\pm B_{\sig\otimes\sig})$ and
$\frac12(1-B_{\tau\otimes\tau})$, and the ensuing relations in the 
three different quotients are, respectively:
\bea (\mathrm{i}) & \Psit^L=\Psit^R,\quad \Psis^L=\Psis^R; \nonumber \\
(\mathrm{ii}) & \Psit^L=\Psit^R,\quad \Psis^L=-\Psis^R; \nonumber \\
(\mathrm{iii}) & \Psit^L=-\Psit^R. \nonumber
\eea
The first case is the trivial boundary; the second the ``fermionic''
boundary where the field $\Psis$ changes sign, and the third the
``dual'' boundary, in which there are two independent fields 
$\Psis^R$ and $\Psis^L$ (corresponding to the order and 
disorder parameter $\sig$ and $\mu$ in Example 1, \sref{E1}). Notice that as 
a representation of either $\B_{\rm 2D}^L$ or $\B_{\rm 2D}^R$ (which are both
isomorphic to the unique maximal local 2D Ising model), the Hilbert 
space \eref{dualhilbert} splits into two inequivalent representations.

\subsection{Classification of boundary conditions}
\label{s:classi}
In this subsection, we shall characterize the minimal central projections,
and hence the irreducible boundary conditions, in a large class of
models: the phase boundaries between two extensions of 
$\A_{\rm 2D}=\A\otimes\A$ that arise as full centres (= $\alpha$-induction
constructions) of (possibly different) chiral extensions.
 
\medskip

The following \tref{t:center} provides a formula for the minimal central 
projections of $\D'\cap \D$. This formula 
allows to compute, in the corresponding irreducible representations, 
the numerical values $\pi_m(\Psi^{L*}_{\sig\otimes\tau}\Psi^R_{\sig\otimes\tau})$, 
i.e., the angles between a pair of charged fields at the boundary, 
without explicitly diagonalizing the algebra \eref{centerF}. In particular, 
explicit knowledge of the coefficients $\zeta$ of the Q-systems is 
not needed. 

In general, not only signs will appear as boundary conditions. E.g.,
in models with a chiral sector of dimension $\dim(\sig)=\gamma=\frac
12(1+\sqrt5)$ (the golden ratio) satisfying the fusion rules
$\sig^2\sim\id\oplus\sig$, it follows from \eref{verlinde} that 
$B_{\sig\otimes\sig}^2 = \gamma^{-2}\cdot\mathbf{1}+\gamma\inv\cdot
B_{\sig\otimes\sig}$. This implies that the spectrum of
$B_{\sig\otimes\sig}$ is $\{1,-\gamma^{-2}=\gamma-2\}$. In the eigenspace 
$B_{\sig\otimes\sig}=\mathbf{1}$ one has $\Psis^L=\Psis^R$, while in the eigenspace
$B_{\sig\otimes\sig}=-\gamma^{-2}$ the two charged fields $\Psis^L$ 
and $\Psis^R$ stand at an angle of $\arccos(-\gamma^{-2})\approx 112.5^\circ$
in the two-dimensional space $H_{\sig\otimes\sig}$ of charged  
fields. 

For \tref{t:center}, we have to assume that $\A_+=\A_-$ and that
its DHR category $\DHR(\A)$ is modular. 
  (It trivially extends to the case when $\DHR(\A_+)$ and $\DHR(\A_-)$ are
  isomorphic modular categories.)
Recall that modularity is
automatic if the chiral theory $\A$ is completely rational \cite{KLM},
and is satisfied for all minimal ($c<1$ Virasoro) models and for many 
current algebra models associated with semisimple Lie algebras
\cite{X} (maybe all), and many other related models \cite{L03,LX}.
This theorem, including the special case \eref{verlinde}, is the only
place where we assume modularity. We also assume that the local
extensions $\B_{\rm 2D}^Y$ ($Y=L,R$) on both  
sides of the boundary are irreducible maximal extensions. These are 
precisely those extensions whose Q-systems are full centres of 
irreducible chiral Q-systems, cf.\ \sref{s:alpha}, thus we assume that
the Q-systems for  
$\A_{\rm 2D}=\A\otimes\A\subset\B_{\rm 2D}^Y$ are given as 
$$(\Theta^L,W^L,X^L) = Z[A],\qquad (\Theta^R,W^R,X^R) = Z[B].
$$ 
The chiral Q-systems $A=(\theta^A,w^A,x^A)$ and $B=(\theta^B,w^B,x^B)$ 
can be different, and they may belong to different Morita equivalence 
classes.  

\medskip

In order to find the minimal projections $I_m$ in
$\Hom(\Theta^R,\Theta^L)$ w.r.t.\ the $\ast$-product, such that 
$E_m=\chi(I_m)$ are the minimal projections in the centre 
$\D_{\rm 2D}'\cap \D_{\rm 2D}$, we have to solve the three equations
\bea\label{projD}
\hbox{self-adjointness:} && I_m^*= F(I_m) \nonumber \\ 
\hbox{idempotency:} && I_m\ast I_{m'} =\delta_{mm'}\cdot I_m, \\
\hbox{completeness:} && \sum_m I_m= W^{L}W^{R*}. \nonumber \eea
Minimality is then automatically ensured if the number
of $I_m$ exhausts the dimension of the centre $\dim(\D_{\rm 2D}'\cap \D_{\rm 2D})=\dim\Hom(\Theta^R,\Theta^L) =
\Tr(Z^{Lt}Z^R)$. Having found the intertwiners $I_m$ corresponding to
the central projections, we obtain the values (= boundary conditions)
$\pi_m(B_\rho)$ from the expansion  
\be\label{WI}
W^L_\rho W^{R*}_\rho = \sum_m \pi_m(B_\rho)\cdot I_m 
\ee
which is the same as \eref{piF} under the linear bijection $\chi$. 

We can now state the main classification result.

\begin{theorem} \label{t:center}
(cf.\ also \cite[Thm.\ 4.44]{BKLR}) Let $\A$ be a completely rational
chiral CFT net, and $\C=\DHR(\A)$ its (modular C* tensor) DHR category. 
Let $\B^Y$ ($Y=L,R$) be two 
local chiral extensions of $\A\otimes\A$ whose commutative Q-systems 
$(\Theta^Y,W^Y,X^Y)$ arise as full centres of chiral Q-systems 
$A=(\theta^A,w^A,w^A)$ and $B=(\theta^B,w^B,w^B)$. Then the 
minimal projections $I_m$ in $[\Hom(\Theta^R,\Theta^L),\ast]$ (i.e., the
solutions to the system \eref{projD}) are given by the operators 
\be\label{ID}
I_{\mm}= \frac{\dim(\beta)}{d_A^2 d_B^2 d_R^2} \cdot
D_{R[\mm]\vert_Z}
\ee
where $\mm=(\beta,m)$ run over the equivalence classes of irreducible
$A$-$B$-bimodules.  

Consequently, $E_\mm = \chi(I_\mm)$ are the minimal central projections in
$\D'\cap \D$ that classify the irreducible boundary conditions
according to \cref{c:bc=mp}.
\end{theorem}

We shall define the operators $D_{R[\mm]\vert_Z}$ below, but we shall
not give a complete proof, which can be found in the review 
\cite[Sect.\ 4.12]{BKLR}. The result is actually implicit in
\cite{FFRS06,KR}, but our proof in \cite{BKLR} is more streamlined 
and benefits from substantial simplifications, 
that apply in the present case of categories of homomorphisms of C* 
algebras, but that are not assumed in \cite{FFRS06,KR}. Namely, to
establish the orthogonality of the projections one has to prove that a
certain operator $Q$ vanishes; because $Q$ is a positive operator, it
is sufficient to show that a faithful positive trace vanishes on $Q$,
cf.\ \cite[p.\ 73]{BKLR}.

Here, we mainly want to explain the notions of the Theorem, and how it
can be used to compute the boundary conditions, i.e., the values 
$\pi_\mm(B_\rho)\in\CC$. 

$A$-$B$-bimodules between Q-systems $A$ and $B$ in a C* tensor category
$\C$ are, in terms of the category, pairs $\mm=(\beta,m)$ where
$\beta$ is an object of $\C$ and $m\in\Hom(\theta^A\beta\theta^B)$
subject to relations, cf.\ \cite{BKLR}. In the case of $\C$ a full subcategory of
$\End_0(N)$, and $A$ and $B$ corresponding to extensions $N\subset M^A$
and $N\subset M^B$ according to \tref{t:vN-Q}, respectively, a
bimodule is equivalent to a homomorphism $\varphi:M^B\to M^A$ such
that $\beta=\ol\iota^A\circ\varphi\circ\iota^A$ and
$m=\ol\iota(v^A\beta(v^B))$. Conversely, every homomorphism 
$\varphi:M^B\to M^A$ corresponds to a bimodule, provided 
$\varphi\prec\ol\iota^A\rho\iota^B$ for some $\rho\in\C$. 
In particular, a Q-system $A$ is an $A$-$A$-bimodule in a natural way 
with $\varphi=\id_M$, hence $m=x^{(2)}\equiv x\cdot x$. 

The $A$-$B$-bimodules for a given pair $A$, $B$ form again a category, 
i.e., one can define intertwiners $\in\Hom(\mm_1,\mm_2)$, with which 
equivalence, inclusion, and direct sums of bimodules can be
defined. $A$-$B$-bimodules can be tensored with $B$-$C$-bimodules 
giving rise to $A$-$C$-bimodules $\mm_{AB}\otimes_B\mm_{BC}$. One
arrives at a (non-strict) 2- category whose objects are the
Q-systems, the 1-morphisms are the bimodules, and 
the 2-morphisms are the bimodule intertwiners. In fact, up to 
unitary equivalence, the notions of inclusion, direct sum, and 
tensor product of bimodules coincide with the corresponding notions of
inclusion, direct sum, and composition in the (strict) 2-category of
homomorphisms as in \sref{s:Q-ext}.   

We now assume that $\C$ is braided, as is the case for $\C=\DHR(\A)$. 
With an $A$-$B$-bimodule $\mm=(\beta,m)$, one can associate an 
intertwiner $D_{\mm}\in\Hom(\theta^B,\theta^A)$ as follows. With
$m\in\Hom(\beta,\theta^A\beta\theta^B)$, one has $\eps_{\theta^A,\beta}
\theta^A\beta(r^{B*})m \in\Hom(\beta\theta^B,\beta\theta^A)$. 
Then the inertwiners
\be\label{Dm}
D_{\mm}:=r_\beta^*\ol\beta
\big(\eps_{\theta^A,\beta}\theta^A\beta(r^{B*})m\big)r_\beta \in 
\Hom(\theta^B,\theta^A)
\ee
with $r_\beta\in\Hom(\id,\ol\beta\beta)$ (part of) a standard solution
to the conjugacy relations for $\beta,\ol\beta$, do not depend on
the standard solution, and depend on $\mm$ only through its
equivalence class as a bimodule. Moreover, these operators form 
a ``representation'' of the tensor category of bimodules in the 
sense that one has $D_{\mm}^* = D_{\ol{\mm}}$ and
\be\label{Dfusion}
D_{\mm_{AB}}D_{\mm_{BC}}=d_B\cdot D_{\mm_{AB}\otimes_B\mm_{BC}} =
d_B\cdot
\sum_{\mm_{AC}}\dim\Hom(\mm_{AC},\mm_{AB}\otimes_B\mm_{BC})\cdot D_{\mm_{AC}}
\ee
where the sum extends over the equivalence classes of irreducible
$A$-$C$-bimodules.

Passing from $\C$ to $\C\boxtimes\C\opp$, a Q-system $A$ in $\C$ trivially
defines a Q-system $A\otimes 1$ in $\C\boxtimes\C\opp$. Similarly,
$A$-$B$-bimodules lift to $A\otimes1$-$B\otimes1$-bimodules. 
The braided product of Q-systems gives rise to a braided product of
bimodules, so that one can define the $R[A]$-$R[B]$-bimodule
$$R[\mm]:=(\mm\otimes 1)\times^+
\mathbf{R},$$ 
where $\mathbf{R}$ is the canonical Q-system $R$ in $\C\boxtimes\C\opp$ 
regarded as an $R$-$R$-bimodule, and $R[A]=(A\otimes 1)\times^+R$. 
The full centre $Z[A]$ is an intermediate Q-system of $R[A]$. One can then
restrict $R[\mm]$ to the full centres $Z[A]$ and $Z[B]$ giving
rise to a $Z[A]$-$Z[B]$-bimodule denoted by $R(\mm)\vert_Z$. 

Now, $D_{R[\mm]\vert_Z}\in\Hom(\Theta^R,\Theta^L)$ in \eref{ID}
is just \eref{Dm} for the $Z[A]$-$Z[B]$-bimodule
$R(\mm)\vert_Z$, where $\mm$ is an irreducible $A$-$B$-bimodule. 

The number of inequivalent irreducible $A$-$B$-bimodules is known
\cite{EP03}. Namely, if $\Theta^L=\bigoplus_{\sig,\tau}
Z^L_{\sig,\tau}\cdot\sig\otimes\tau$ and $\Theta^R=\bigoplus_{\sig,\tau}
Z^R_{\sig,\tau}\cdot\sig\otimes\tau$, then this number is 
$\Tr(Z^{Lt}Z^R)$ which equals the dimension of the centre 
$\dim\Hom(\Theta^R,\Theta^L)$. Thus, provided $D_{R[\mm]\vert_Z}$ are
linearly independent, they form a basis of $\Hom(\Theta^R,\Theta^L)$. 

Using the non-degeneracy property of the braiding in a modular
category, one proves that they are indeed linearly independent, and in fact orthogonal w.r.t.\ the scalar product 
\be\label{scp}
(D_1,D_2):= R^{R*}\cdot D_1^*D_2\cdot R^R = R^{L*}\cdot D_2D_1^*\cdot
R^L,
\ee
namely (e.g., \cite[Sect.\ 4.12]{BKLR})
$$(D_{R[\mm_1]\vert_Z},D_{R[\mm_2]\vert_Z}) =
d_A^2d_B^2d_R^4 \cdot \delta_{\mm_1,\mm_2}.$$
Using this fact together with the observation that the scalar product
\eref{scp} is related to the $\ast$-product by
$(D_1,D_2) = W^{L*}\cdot(F(D_1^*)\ast D_2)\cdot W^R$,
we can also prove \cite{BKLR} that, up to normalizations as given in 
\eref{ID}, $D_{R[\mm]\vert_Z}$ solve \eref{projD}. 

\medskip 

As an interesting side-result one obtains \cite{KR,BKLR}

\begin{proposition}
\label{p:FullCentreDim}
All full centre Q-systems $Z[A]$ ($A$ an irreducible Q-system 
in a modular tensor category $\C$) have the same dimension
$$d_{Z[A]}^2\equiv\dim(\Theta^{Z[A]}) \equiv\sum_{\sig,\tau}
Z_{\sig,\tau}\dim(\sig)\dim(\tau) = d_R^2 = \sum_\rho\dim(\rho)^2.$$ 
where the sum extends over the equivalence classes of irreducible
objects of $\C$.  
\end{proposition}

\subsection{Computation of the boundary conditions}
\label{s:comp}
Now, we turn to computing the values $\pi_m(B_\rho)$. W.r.t.\ the
scalar product \eref{scp}, also the basis $T:=W^L_\rho W^{R*}_\rho$ of
$\Hom(\Theta^R,\Theta^L)$ (with $W^Y_\rho$ normalized as in
\sref{s:bcond}) is orthogonal:  
$$(T,T') = \frac{d_R^2}{\dim(\rho)} \cdot \delta_{T,T'}$$ 
(where the label $T$ is understood to include the data $\rho$ and
possible multiplicity indices). Therefore, the
passage from the normalized basis $\frac1{d_Ad_Bd_R^2}\cdot
D_{R[\mm]\vert_Z}$ to the normalized basis
$\frac{\sqrt{\dim(\beta)}}{d_R}\cdot T$ defines a unitary matrix  
\be\label{SmT}
S^{AB}_{\mm,T}:= \frac{\sqrt{\dim(\rho)}}{d_Ad_Bd_R^3}\cdot
(D_{R[\mm]\vert_Z},T),
\ee
and hence 
$$W^L_\rho W^{R*}_\rho \equiv T = 
\frac{d_R}{\sqrt{\dim(\rho)}}\sum_{\mm} S^{AB}_{\mm,T} \cdot
\frac1{d_Ad_Bd_R^2}\cdot D_{R[\mm]\vert_Z} =
\frac{d_Ad_Bd_R}{\sqrt{\dim(\rho)}}\sum_{\mm} S^{AB}_{\mm,T}
\cdot\frac1{\dim(\beta)}\cdot I_{\mm}.$$ 
This is the desired decomposition \eref{WI}, i.e., the central
operators $B_\rho=\Psi^{L*}_\rho \Psi^R_\rho$ take the values
\be\label{piBS}
\pi_{\mm}(B_\rho)=
\frac{d_Ad_Bd_R}{\dim(\beta)\sqrt{\dim(\rho)}}\cdot S^{AB}_{\mm,T} \qquad
(\mm = (\beta,m),\; T=W^L_\rho W^{R*}_\rho).
\ee

\begin{corollary} \label{c:bound=bim}
The irreducible boundary conditions, i.e., the irreducible
representations of $\D=\B^L_{\rm 2D}\times^-\B_{\rm 2D}^R$ are labelled by the
chiral bimodules between the chiral Q-systems whose full centres give
rise to the extensions $\B_{\rm 2D}^L$ and $\B_{\rm 2D}^R$. The ``angles''
$\pi_{\mm}(\Psi^{L*}_\rho\Psi^R_\rho)$ between left and right fields 
are given by the matrix elements of the generalized $S$-matrix as in \eref{piBS}. 
\end{corollary}

The remaining task is to find an efficient way to compute the 
matrix elements $S^{AB}_{\mm,\rho}$. First, we list some rather trivial 
special cases.

We have 
$$S^{AB}_{\mm,\id}= \frac{\dim(\beta)}{d_Ad_Bd_R}$$
for $T=W^LW^{R*}$ the unit w.r.t.\ the $\ast$-product. This just
reflects the fact that $\Psi^L_\id=\Psi^R_\id=\mathbf{1}$, hence the
corresponding operator $B_\id$ must be $=\mathbf{1}$ in every representation. 

For $A=B$ (i.e., isomorphic QFT on both sides of the boundary), and
$\mm=A$ as an $A$-$A$-module, one finds $D_{R(A)\vert_Z} = d_A^2d_R\cdot
1$, hence  
$$S^{AA}_{A,\rho} = \frac{\sqrt{\dim(\rho)}}{d_R}.$$
Consequently, $\pi_A(B_\rho)=\mathbf{1}$ and hence $\Psi^L_\rho=\Psi^R_\rho$
for all $\rho$. Thus, the trivial bimodule describes the trivial boundary. 

For $A=B=(\id,1,1)$ the Q-system of the trivial chiral extension
$\A\subset\A$, the bimodules are just the irreducible chiral sectors 
$\sig\in\C$. Since $Z[A]=R$ is the canonical Q-system, $\rho^+\otimes\rho^-$ are of the form
$\rho\otimes\ol\rho$, and one recovers the example
\eref{verlinde}
$$S^{11}_{\sig,\rho\otimes\ol\rho} =S_{\sig,\rho}.$$

\medskip

One may use further special properties of the matrix \eref{SmT} in
order to compute its matrix elements more efficiently than by evaluation of the
defining scalar products. Namely, the property \eref{Dfusion} lifts
to $R[\mm]\vert_Z$: 
$$ D_{R[\mm_{AB}]\vert_Z} D_{R[\mm_{BC}]\vert_Z} = d_Bd_R^2\cdot
D_{R[\mm_{AB}\otimes_B\mm_{BC}]\vert_Z} = d_Bd_R^2\cdot
\sum_{\mm_{AC}} N_{\mm_{AB},\mm_{BC}}^{\mm_{AC}} \cdot D_{R[\mm_{AC}]\vert_Z},$$
where we write the fusion rules of irreducible $A$-$B$-bimodules and
$B$-$C$-bimodules as 
$$\mm_1\otimes_B\mm_2 \simeq \bigoplus_{\mm_3}
N_{\mm_1,\mm_2}^{\mm_3}\cdot \mm_3.$$
If we now express $D_{R[\mm]\vert_Z}$ in terms of the bases $T=W^L_\rho
W^{R*}_\rho$, we arrive at
$$\sum_{T_1,T_2} S^{AB}_{\mm_1,T_1} S^{BC}_{\mm_2,T_2}\cdot
\sqrt{\dim(\rho)} \cdot
T_1T_2 = \sum_{\mm_3} N_{\mm_1,\mm_2}^{\mm_3}\sum_{T_3} S^{AC}_{\mm_3,T_3}\cdot T_3 ,$$
which can be solved for the fusion rules: 
\be\label{genverl}
N_{\mm_1,\mm_2}^{\mm_3} =
\sum_{T_1,T_2,T_3}\frac{\dim(\rho)^{3/2}}{d_R^2} (T_3,T_1T_2)\cdot
S^{AB}_{\mm_1,T_1} S^{BC}_{\mm_2,T_2}\ol{S^{AC}_{\mm_3,T_3}}.
\ee
If there are no multiplicities (i.e., $\dim\Hom(\rho,\Theta^Y)= 0$ or
$1$ for all $Y=A,B,C$), then $(T_3,T_1T_2)=d_R^3\dim(\rho)^{-2}\cdot
\delta_{T_1,T_3}\delta_{T_2,T_3}$, and \eref{genverl} simplifies to
the ``generalized Verlinde formula''
\be\label{specverl}
N_{\mm_1,\mm_2}^{\mm_3} =
\sum_{T}\frac{d_R}{\sqrt{\dim(\rho)}} \cdot
S^{AB}_{\mm_1,T} S^{BC}_{\mm_2,T}\ol{S^{AC}_{\mm_3,T}} .
\ee
In particular, for $B=C$, this becomes
\be\label{verl}
N_{\mm_1,\mm_2}^{\mm_3} =
\sum_{T}
\frac{S^{AB}_{\mm_1,T} S^{BB}_{\mm_2,T}\ol{S^{AB}_{\mm_3,T}}}{S^{BB}_{\mathbf{B},T}},
\ee
and the unitary matrices $S^{AB}_{\mm,T}$ diagonalize the
fusion rules:
\be\label{fusdiag} \sum_{\mm_1,\mm_3}\ol{S^{AB}_{\mm_1,T}} N_{\mm_1,\mm}^{\mm_3}
S^{AB}_{\mm_3,T'}=\frac{S^{BB}_{\mm,T}}{S^{BB}_{\mathbf{B},T}}\cdot \delta_{T,T'}.
\ee
Because fusion rules are comparatively easy to compute with the help of
Frobenius reciprocity, one may use \eref{fusdiag} to determine the
matrices $S^{AB}$ (with some complication to be expected if there are
multiplicities). 

Together with \eref{piBS}, this completes the computation of the boundary
conditions. 

\medskip

We want to stress that in this way, the boundary conditions are
ultimately characterized by chiral data only, namely the fusion
category of chiral bimodules, which are in turn the irreducible
subobjects of $\iota_L\circ\rho\circ\ol\iota_R$, $\rho\in\C$. 

\subsection{Example 6: Chiral boundary fields in the Ising model}
\label{E6}

In the Ising model, the explicit computations of the algebra of
charged operators at a phase boundary are sufficiently simple to be
carried out by hand, by using the defining relations of the braided product
Q-system and the relations of the Ising model given in the previous
Examples 2--5, \sref{E2}, \sref{E3}, \sref{E4}, \sref{E5}.

Let us consider the operators
\be\label{chiral}
\psi_+ := \sqrt{-i}\cdot\Psis^{L*}(u\otimes 1)\Psis^R,\qquad \psi_- :=
\sqrt{i}\cdot\Psis^{L*}(1\otimes u)\Psis^R,
\ee
which are by construction chiral intertwiners: 
$$\psi_+ a = (\tau\otimes \id)(a)\psi_+,\qquad \psi_- a = (\id\otimes
\tau)(a)\psi_- \qquad (a\in \A_{\rm 2D}).$$
Thus, they belong to the relative commutants $(\mathbf{1}\otimes \A)'\cap
\D_{\rm 2D}$ and $(\A\otimes \mathbf{1})'\cap \D_{\rm 2D}$, respectively, and in fact 
the latter are generated by $(\A\otimes \mathbf{1})\vee\psi_+$ 
and $(\mathbf{1} \otimes \A)\vee\psi_-$, respectively. They satisfy 
the relations 
$$\psi_\pm^* = \psi_\pm,\qquad \psi_\pm^2 =
\frac12(1-\Psit^L\Psit^R),\qquad \psi_+\psi_- = -\psi_-\psi_+=\frac i2(\Psit^R-\Psit^L).$$
Thus, in the representations (i) and (ii) in \sref{E5}, $\psi_\pm=0$, while in the
``dual'' representation (iii), they are a pair of left- and right-moving
selfadjoint unitary Fermi fields implementing the chiral automorphisms $\tau$, 
such that 
$$\psi_+\psi_-=-\psi_-\psi_+=i\Psit^R=-i\Psit^L.$$
The operator $C_\sig:=\Psis^{L*}(u\otimes
  u)\Psis^R=\Psis^{L*}\Psis^R\Psit^R$ (by \sref{E4}) is equal to
  $\pm\Psit$ in (i) and (ii), and vanishes in (iii). The operator
  product between $\Psis^L$ and $\Psis^R$ is computed as 
\bea\label{ope}
\sqrt2 \cdot \Psis^L\Psis^R = (r\otimes r) B_\sig +(t\otimes t)C_\sig 
+ \sqrt{i}\cdot(t\otimes r)\psi_+ +\sqrt{-i}\cdot(r\otimes t)\psi_-,
\eea
which is $=\pm (r\otimes r)\mathbf{1} + (t\otimes t) \Psit$ in (i) and (ii),
and $=\sqrt{i}\cdot(t\otimes r)\psi_+ +\sqrt{-i}\cdot(r\otimes t)\psi_-$ in
(iii). 

We also compute, e.g., $\sqrt{i}\cdot\psi_+\Psis^R=
\frac12(\Psis^L-\Psit^R\Psis^L)$ which again vanishes in (i) and (ii)
(because $\Psit^R=\Psit^L$ and $\Psit^L\Psis^L=\Psis^L$),
and equals $\Psis^L$ in (iii). We obtain
\be\label{LchiR}
\Psis^L=\sqrt{\pm i}\cdot \psi_\pm\Psis^R,\qquad\Psis^R= \sqrt{\mp
  i}\cdot \psi_\pm\Psis^L
\ee
(valid in (iii)), thus expressing the left fields in terms of the
right fields and the Fermi fields, and vice versa $R\leftrightarrow
L$. This is the AQFT analogue of \eref{dualF} and \eref{dualF2}, when 
$\Psis^R$ is identified with the order parameter $\sig$ and $\Psis^L$ 
with the disorder parameter $\mu$.

\medskip

It would be interesting to understand in the general case, whether and
how chiral (nonlocal) fields emerge as certain products of fields from
$\B^L_{\rm 2D}$ and from $\B^R_{\rm 2D}$ (i.e., as a subalgebra of a representation 
of the nonlocal braided product), as exemplified in \eref{chiral}, \eref{ope}, and 
how, conversely, the local fields of $\B^L_{\rm 2D}$ can be represented as
certain products of chiral fields and local fields from $\B^R_{\rm 2D}$ (and
vice versa), as exemplified by \eref{LchiR}. 

\subsection{Classification of factorial defects for maximal local extensions}

Recall from \cite{LR04} that full centre extensions are
maximal local extensions. Hence, in view of \cref{c:B<Dloc}, for a
defect between two full centre extensions $\B^L$, $\B^R$, one has 
$$\B^L=\D\loc^-\quad\hbox{and}\quad\B^R= \D\loc^+.$$
This implies

\begin{lemma}
\label{l:morita}
Let $\B^Y$ ($Y=L,R$) be two 
local chiral extensions of $\A\otimes\A$ whose commutative Q-systems 
$(\Theta^Y,W^Y,X^Y)$ arise as full centres of chiral extensions.

Then all (not necessarily irreducible but) factorial
defects between $\B^L$ and $\B^R$, i.e., factorial extensions
$\D\supset \A\otimes \A$ such that $\B^L = \D\loc^-$
and $\B^R = \D\loc^+$, are Morita equivalent. \end{lemma}

\begin{proof}
Let us apply the full centre construction (again) to $\A\otimes
\A\subset \D$ (for example we can see it as a time $t=0$ net) to get
an extension $(\A\otimes \A)\otimes (\A\otimes \A)\subset \D^{(2)}$
given by a commutative Q-system $Z[D]$ in
$(\C\boxtimes\C\opp)\boxtimes  (\C\boxtimes\C\opp)\opp$. 

Applying \eref{eq:MaximalChiral} to this construction, it follows that 
\bea\label{double}
(\A\otimes \A)\otimes(\A \otimes \A)\subset \B^R\otimes \B^L\subset
\D^{(2)}.
\eea 
But the dimension of $(\A\otimes \A)\otimes(\A \otimes \A)\subset \B^R\otimes \B^L$ 
by \pref{p:FullCentreDim} is
$$
d^2_{B^L\otimes B^R}
=\Big(\sum_\rho\dim(\rho)^2\Big)^2=
\sum_{\rho,\sigma} \dim(\rho\otimes\bar\sigma)^2 = d^2_{Z[D]}\,,
$$
thus coincides with the dimension of $Z[D]$ by using
\pref{p:FullCentreDim} again. But this means we have 
$\B^R\otimes \B^L= \D^{(2)}$.
Because two extensions are Morita equivalent if and only if they have
isomorphic full centres \cite{KR} all extensions $\A\otimes\A\subset
\D$ with the stated property are Morita equivalent.  
\end{proof}

\begin{remark}\label{r:T} 
Let $\B^Y$ ($Y=L,R$) be two 
local chiral extensions of $\A\otimes\A$ whose commutative Q-systems 
$A^Y$ arise as full centres of chiral Q-systems. 
The main result Prop.\ 4.3 in \cite{KR} is obtained by applying first the full
centre and then the adjoint functor $T$. If we follow their strategy
and apply the full centre to the Q-system of the extension
$\A\otimes\A\subset \D$ and then apply $T$, it is easy to check that
we obtain the  Q-system $A^L\otimes^- A^R\cong A^R\otimes^+A^L$. 
\end{remark}

Together with the previous \lref{l:morita} this gives a
classification of irreducible defects as in \dref{d:bc}.

\begin{theorem}
Let $\B^Y$ ($Y=L,R$) be two local chiral extensions of $\A\otimes\A$
whose commutative Q-systems $(\Theta^Y,W^Y,X^Y)$ arise as full centres
of chiral Q-systems $A^L=Z[A]$ and $A^R=Z[B]$.

Then there is a one-to-one correspondence between factorial defects
$$(\D\loc^-=)\, \B^L\subset\D\supset \B^R \, (=\D\loc^+).$$
and  equivalence classes of (not necessarily irreducible) $A$-$B$-bimodules. 
The defect is a boundary condition if and only if
  the bimodule is irreducible. In this case, the boundary
  condition is irreducible.
\end{theorem}

\begin{proof}
Because the index is finite, the relative commutant is finite. We can
decompose the Q-system giving $\A\otimes \A\subset \D$ into irreducible
Q-systems. We first claim that each irreducible component is again a
defect. Indeed, if we take an irreducible component $\A \subset \D_i$,
then $\A\subset \D$ and $\A\subset \D_i$ are Morita equivalent by
\tref{t:moritadirectsum} and 
\lref{l:morita}, hence they have the same full centre in 
$(\C\boxtimes\C\opp)\boxtimes (\C\boxtimes\C\opp)\opp$, and hence
their maximal local subalgebras $(\D_i)^+\loc\otimes(\D_i)^-\loc$
according to \eref{eq:MaximalChiral} coincide with
$\D^+\loc\otimes\D^-\loc = \B^R\otimes\B^L$. In particular, each
$\D_i$ contains $\B^L\vee\B^R$, and hence is a defect, which is obviously
irreducible. By \rref{r:T}, $\D_i$ is inside the central decomposition of the
universal construction \pref{p:universal}, and thus an irreducible boundary
condition as classified in \tref{t:center}.
Conversely, for every $A$-$B$-bimodule $\mm$ we decompose it as a direct sum 
$$
\mm=\bigoplus n_i \mm_i
$$
of irreducible ones and obtain a factorial boundary condition by taking a
corresponding ``direct sum'' as in \tref{t:moritadirectsum} (with the
same multiplicities) of the extensions $\A\otimes \A\subset
\D_i$. Notice that \tref{t:moritadirectsum} applies thanks to \lref{l:morita}.

We have already shown that if the bimodule is irreducible, that the
defect is indeed an irreducible boundary condition.
So let $\bigoplus_i n_i \mm_i$ be a reducible bimodule. We want to show that the
corresponding defect $A\subset D$ is not a boundary condition,
i.e., $\B^L\vee \B^R \neq \D$. 
But in this case we have intermediate inclusions
$\A \subset \bigoplus \Mat_{n_i}(\B^Y) \subset
\bigoplus \Mat_{n_i}(\D_i) \subset \D$ for $Y=L,R$,
thus $\B^L \vee \B^R$ can generate at most $\bigoplus
\Mat_{n_i}(\D_i)\subsetneq \D$. (They will actually generate only the
diagonal subalgebras of the matrix algebras.) 
Thus $\D$ is only a defect and not a boundary condition.
\end{proof}

\section{Outlook: boundaries and local gauge transformations}
\setcounter{equation}{0}
\subsection{Spacelike boundaries?}\label{s:spacelike}
One could attempt to repeat the same analysis for spacelike boundaries, 
i.e., $t=0$ lines in some Lorentz system, separating local QFTs $\B_{\rm 2D}^F$ 
(defined in the future of the boundary) and $\B_{\rm 2D}^P$ (defined in the past) 
with a common subtheory including the stress tensor. Again, both nets 
extend to the entire Minkowski spacetime. 

But in this case, causality requires that $B^P_{\rm 2D}$ and $B^F_{\rm 2D}$ must be 
mutually local, whereas the ``universal construction'' 
$B^P_{\rm 2D}\times^\pm B^F_{\rm 2D}$ will in general not be local. An easy exercise 
shows that it is local if and only if $\eps_{\Theta^F,\Theta^P}^+ = 
\eps_{\Theta^F,\Theta^P}^-$. 

On the other hand, the braided product may become local in some
representations, e.g., if $B^P_{\rm 2D}=B^F_{\rm 2D}$ this happens for
the central projection induced by the trivial bimodule, giving rise to
a quotient in which $\Psi^F_\rho=\Psi^P_\rho$, as in \sref{s:comp}. It
would be an interesting question to characterize all central
projections which give rise to a local net.

\subsection{Four spacetime dimensions}
Let us try and draw some lessons for four-dimensional QFT, although
the situation departs from the present setting in several
respects. This leads us to some speculative considerations concerning
local gauge transformations in algebraic QFT. 

In four dimensions, there is no distinction between ``left'' and
``right'', and the braiding is in fact a symmetry, 
i.e., $\eps^+_{\sig,\rho}=\eps^-_{\sig,\rho}$. Consequently, 
there is only one braided product, and the braided product of 
two local extensions of a net of observables is indeed local, namely 
the commutation relations
$$\Psi^1_\rho\Psi^2_\sig = \eps_{\sig,\rho} \cdot
\Psi^2_\sig\Psi^1_\rho$$
derived from the braided product imply that the generating charged
fields $\Psi^1_\rho$ and $\Psi^2_\sig$ commute whenever $\rho$ and
$\sig$ are localized at spacelike distance. 

Causality at a timelike boundary requires ``left locality'' as before. 
But left locality implies mutual locality of the QFTs on either side of 
the boundary, and the braided product is again a universal construction. 
One may as well admit spacelike boundaries, since (unlike 
in two dimensions, cf.\ \sref{s:spacelike}) they do
not require stronger locality properties than timelike boundaries.

Indeed, the original development of the DHR theory of superselection 
sectors was aimed at physical spacetime, and conformal symmetry was not
assumed. Also in our previous analysis, conformal symmetry was not
directly used, it only served to guarantee the existence of chiral
observables so that the structure $\DHR(\A_+\otimes\A_-) =
\DHR(\A_+)\boxtimes \DHR(\A_-)\opp$ arises which became essential for the
classification of boundary conditions in \sref{s:classi}. 

In general the tensor category $\DHR(\A)$ will not be
rational. Its irreducible objects are given by the unitary
representations of an intrinsically determined compact gauge group $G$,
and there is a distinguished graded local extension $\A\subset\B$
(the Doplicher-Roberts field algebra \cite{DR}) with a faithful action 
of $G$ by automorphisms (global gauge transformations) 
such that $\A=\B^G$. Since the index of $\A\subset\B$ equals
the order of the group $G$, this extension will not be described by a
Q-system proper, unless $G$ is finite. One would have to adapt
the theory to Q-systems for extensions of infinite index with suitable 
regularity properties. Let us for the sake of this outlook assume $G$
finite. The distinguished DR extension is maximal in the sense that
every irreducible extension is intermediate to $\A\subset \B$, and is
given by $\B^H$ with $H\subset G$ some subgroup of $G$. 

For the braided product of the DR extension $\B$ with
itself, $\Hom(\Theta,\Theta)$ is isomorphic to the group algebra $\CC
G$. Let us assume for the moment that $\B$ is local, i.e., $\A$ has no 
fermionic sectors. Then the centre of the braided product 
$\B\times\B$ is given by $\Hom(\Theta,\Theta)$ w.r.t.\ the 
$\ast$-product, which is isomorphic to the algebra $C(G)$ of
functions on the group, and the minimal projections are the
$\delta$-functions $\delta_g\in C(G)$, cf.\ \cite{WH}. 
Hence, the boundary conditions are labelled by the elements $g\in G$, 
and the relations between the fields on both sides of the boundary 
are given by 
$$\pi_g(\Psi^2_\rho)= \pi_g(\alpha_g(\Psi^1_\rho)) = u_\rho(g\inv)\cdot\pi_g(\Psi^1_\rho)$$
with $u_\rho$ being a unitary representation of $G$ acting in the
$\dim(\rho)$-dimensional space of charged intertwiners
$H_\rho=\Hom(\iota,\iota\rho)\subset\B$. 

Thus, a boundary can be regarded as a vector bundle (with a two-point
basis and the charged fields as fibre) in which a boundary condition 
defines a parallel transport by a gauge transformation. One may now 
imagine a lattice in Minkowski spacetime with spacelike and timelike 
edges, and its dual subdivision of spacetime into cells with 
timelike and spacelike faces as boundaries. The corresponding multiple 
braided product of field algebras has a huge centre (one copy of
$C(G)$ for each edge), and its minimal central projections are gauge 
configurations on the lattice. We suggest that this is the kinematical
arena in which one should place algebraic QFT theory with local gauge
symmetry.   

Of course, several technical aspects have to be addressed. First, the
braided product of extensions must be defined when $G$ is of infinite
order, such that the centre becomes $C(G)$. Second, the faces intersect
in surfaces of co-dimension 2, thus there will arise boundary conditions 
of higher order. Third, the case of fermionic sectors has to be 
included, such as to cover cases like Quantum Electrodynamics 
where $\A$ is to be thought of as the electrically neutral subalgebra 
of the Maxwell-Dirac algebra. The centre of the braided product of two 
graded-local extensions will be smaller than $C(G)$; we expect 
it to be $C(G/\ZZ_2)$ where $\ZZ_2\subset G$ is the grading 
automorphism. Furthermore, a continuum limit has to be devised in
which local algebras arise by refining the cells in a given region. 

The most important element, however, still missing in our suggestion,
is a formulation of the ``gauge dynamics'', i.e., the dynamics of the
gauge field itself and its ``minimal coupling'' to the gauge covariant
fields. In the picture above, the infinite braided product algebra is
just a direct sum of algebras, one for each gauge configuration.

Gauge curvature is described in terms of closed paths (plaquettes) 
in the lattice, which necessarily come along with intersecting faces. 
It may be speculated that the associated boundary conditions of higher
order carry additional degrees of freedom that play a dynamical
role. Adding these degrees of freedom should embed the infinite
braided product with its huge centre into a larger field algebra that
may even be irreducible.  

We hope to come to a better understanding of these (presently highly
speculative) issues in future work. 

\bigskip

{\large\bf Acknowledgments} 

\medskip

We are very much indebted to J. Fuchs, I. Runkel, and C. Schweigert 
for their hospitality and their patience in explaining us their
work. R.L. also thanks J. Fr\"ohlich and F. Xu for stimulating 
motivational comments. M.B. thanks A. Henriques for his hospitality
and explaining their approach \cite{BDH} to defects of conformal
nets. We finally thank the referee for a number of critical questions
and valuable comments which helped to correct a few errors and led to
a clearer exposition of several issues.

\end{document}